\documentclass[seceq]{ptptex}
\usepackage{graphicx}
\usepackage{here}
\usepackage{wrapfig}
\usepackage{wrapft}
\newcommand{\beq}{\begin{eqnarray}}
\newcommand{\eeq}{\end{eqnarray}}
\newcommand{\la}{\langle}
\newcommand{\ra}{\rangle}

\newcommand{\bk}{\mbox{\boldmath $k$}}
\newcommand{\br}{\mbox{\boldmath $r$}}
\newcommand{\bx}{\mbox{\boldmath $x$}}
\newcommand{\by}{\mbox{\boldmath $y$}}

\newcommand{\bL}{\mbox{\boldmath $L$}}

\newcommand{\bS}{\mbox{\boldmath $S$}}

\newcommand{\bsigma}{\mbox{\boldmath $\sigma$}}

\newcommand{\brs}{\mbox{\scriptsize \boldmath $r$}}
\newcommand{\bxs}{\mbox{\scriptsize \boldmath $x$}}

\newcommand{\pslash}{p\kern-1ex /}
\newcommand{\kslash}{k\kern-1ex /}
\newcommand{\qslash}{q\kern-1ex /}
\newcommand{\lslash}{l\kern-1ex /}
\newcommand{\sslash}{s\kern-1ex /}
\newcommand{\paslash}{p_a\kern-2ex /}
\newcommand{\pbslash}{p_b\kern-2ex /}
\newcommand{\Dslash}{{\cal D}\kern-1.5ex /}
\newcommand{\dslash}{\partial\kern-1.2ex /}

\newcommand{\Eq}[1]{Eq.~(\ref{#1})}
\newcommand{\Fig}[1]{Fig.~\ref{#1}}
\newcommand{\Table}[1]{Table~\ref{#1}}

\usepackage[dvips]{color}

\title{
Nucleon-Nucleon Potential and its Non-locality in Lattice QCD
}

\author{
  Keiko \textsc{Murano}$^{1}$$\footnote{address after April 1st, 2011:
  RIKEN Nishina Center, RIKEN, Wako 351-0198, Japan}$,
  Noriyoshi \textsc{Ishii}$^{2}$,
  Sinya \textsc{Aoki}$^{2,3}$,
  Tetsuo \textsc{Hatsuda}$^{4,5}$
}
\inst{
$^1$KEK Theory Center, Institute of Particle and Nuclear Studies,\\
High Energy Accelerator Research Organization (KEK), \\
Tsukuba, Ibaraki 305-0801, Japan\\
$^2$Graduate School of Pure and Applied Sciences, University of Tsukuba,\\
 Tsukuba, Ibaraki 305-8571, Japan\\
$^3$Center for Computational Sciences, University of Tsukuba, \\
Tsukuba, Ibaraki 305-8577, Japan\\
$^4$Department of Physics, The University of Tokyo, Tokyo 113-0033, Japan\\
$^5$Institute for the Physics and Mathematics of the Universe (IPMU),\\
The University of Tokyo, Chiba 277-8583, Japan
}
\abst{  
  By the quenched  lattice QCD simulation for two nucleons
  with finite scattering energy,
   validity of the derivative expansion
 of the general nucleon-nucleon potential 
  $ U(\br,\br')=  V(\br, {\nabla}_{\brs})  \delta^3(\br-\br^\prime)$ is studied.  
  The relative kinetic energy between two 
   nucleons is introduced through the  anti-periodic boundary condition
 in the spatial directions.  On a hypercubic lattice with the 
 lattice spacing $a\simeq 0.137$ fm and the spatial extent $L_{\rm s} \simeq 4.4$ fm
  with the pion mass  $m_{\pi}\simeq 530$  MeV, the local 
  potentials for two different energies ($E \simeq  0$ MeV and $45$ MeV) are compared
  and found to be identical within statistical errors, 
   which validates the 
local approximation
 of  $U(\br,\br')$ up to $E=45$ MeV 
  for the central and tensor potentials.  
  Central potentials in the spin-singlet channel
  for different orbital angular momentums ($\ell=0$ and $\ell=2$)
  at $E \simeq 45$ MeV are also found to be the same   within the errors,
  which also supports the local approximation.
}

\begin{document}

\maketitle

\section{Introduction}
 The nucleon-nucleon  (NN) potential \cite{Machleidt:2000ge,Wiringa:1994wb,Stoks:1994wp}
 is a fundamental quantity to
 study various properties of atomic nuclei and nuclear matter.
 Recently,  a first attempt to calculate
  the NN potential from QCD was reported on the 
 basis of the Nambu-Bethe-Salpeter (NBS) wave function for the 
  two nucleons on the lattice 
  \cite{Ishii:2006ec,Aoki:2009ji}. Also, the method has been  
  extended to the baryon-baryon (BB) interactions with strangeness 
 \cite{Nemura:2008sp,Inoue:2010hs,Inoue:2010es},
 the three-nucleon interaction \cite{Doi:2010yh} and  meson-baryon interactions
\cite{Ikeda:2010sg,Kawanai:2010ev}. 
 Since the NN interaction is short ranged, the NN potential 
  extracted from lattice QCD simulations is exponentially insensitive to the 
   spatial lattice extent $L_{\rm  s}$ as long as  $L_{\rm  s} \gg 1/m_{\pi}$. Then
 one can calculate observables such as the 
  scattering phase shifts by employing the lattice NN potential and 
  solving the Schr\"{o}dinger equation in the infinite volume. 

 In general, the lattice NN potential obtained from the 
 NBS wave function is  energy-independent but 
  non-local, $U(\br,\br')$.
 In practice,  $U$ is  rewritten in terms of an infinite set of energy-independent 
 local potentials $V^{(\rm{LO})}(\br)$, $V^{(\rm{NLO})}(\br), \cdots$, 
 by the derivative expansion. 
 These local potentials are determined successively by measuring the NBS wave functions
  for different scattering energies $E$ below the inelastic threshold $E_{\rm th}$.
  A possible criterion for the validity of the derivative expansion at low energies 
  is  the stability of the local potentials 
  against the variation of 
  the scattering energy in the interval $0 \le E < E_{\rm th}$.
  \footnote{Note that $E_{\rm th}$ is an observable determined by the 
   pion mass and is considerably smaller than the scale of the lattice cutoff 
   $a^{-1}$.
}

  The purpose of this paper is to check such stability through the lattice data at 
  $E\simeq 0$ MeV and  $E\simeq 45$ MeV: these two cases are
   realized on the lattice by taking the 
   periodic and anti-periodic boundary conditions in the spatial directions. 
 We carry out  
 quenched lattice QCD simulations  with $L_{\rm s}\simeq 4$ fm and the pion mass
  $m_{\pi}\simeq 530$  MeV.  We will show that
  the leading-order local potentials at the above two different energies  show
  no difference within statistical  error,
   which validates the
local approximation up to $E=45$ MeV 
  for the central and tensor potentials.
  \footnote{In Ising field theory, it is analytically shown that the
  energy-dependence is weak at low energy, indicating that the
  non-locality of the potential is weak \cite{Aoki:2008yw}.}
Difference of the spin-singlet central
   potentials between $\ell=0$ and $\ell=2$  is also studied, with $\ell$ being the orbital angular momentum.
 A preliminary account of these results is given in Refs.\citen{Murano:2010hh,Murano:2010tc}.
  
  This paper is organized as follows.
 In Sec.\ref{V-intro}, we make a brief review on the energy-independent
 non-local potential and its derivative expansion. An explicit construction
  of the leading order terms of the derivative expansion is also presented.
 In Sec.\ref{sec:boundary}, we explain a method to realize non-zero energy NN scattering
  on the lattice through the spatial boundary conditions. In particular, we introduce 
  a novel momentum wall source operators which are suitable for the purpose of the 
   present paper.
 In Sec.\ref{sec:numerical}, we present numerical results for the NBS wave functions and 
  the associated leading order potentials for different $E$ and $\ell$.
 Sec.\ref{summary} is devoted to summary and concluding remarks.
In Appendix \ref{sec:lattice_sym}, we give a brief summary of the representation of the 
cubic group used in this paper.  
In Appendix \ref{sec:D-wave-op}, some details of constructing the $\ell=2$ source
 operator by using the cubic group representation is presented.

\section{Non-local NN potential and its derivative expansion}
\label{V-intro}
To  define  the  NN  potential  in QCD,  we  consider  the  equal-time
Nambu-Bethe-Salpeter (NBS)  wave function in  the center of  mass (CM)
frame defined by
\begin{eqnarray}
  \phi_{\alpha\beta}(\br;k)
  &\equiv&
  \la 0 |
  p_\alpha(\bx)
  n_\beta(\by)
  | B=2;k \ra ,
  \hspace{3em}
  (\br
  \equiv
  \bx-\by), \label{eq:BSwave}
\end{eqnarray}
where $\vert B=2;k\rangle$ is a QCD eigenstate with baryon
number two ($B=2$), and $p_{\alpha}(x)$,  $n_{\beta}(y)$  are local  composite  nucleon
operators with spinor indices $\alpha$ and $\beta$. 
The asymptotic relative momentum $k$  is related to the  relativistic total energy
$W$ as $W=2\sqrt{m^2_N+k^2}$ with $m_N$ being the nucleon mass.
In the following, we  consider  the  elastic  region where
$W  <  W_{\rm th}\equiv 2 m_N + m_\pi$ is satisfied with the pion mass $m_\pi$.

The asymptotic behavior of the NBS wave function for $|\br| > R$
 ($R$ being the typical interaction range) 
  is characterized by the scattering  phase shift for hadrons  
  \cite{Luscher:1986pf,Luscher:1990ux,Lin:2001fi,Aoki:2005uf,Aoki:2009ji,Ishizuka:2009bx}.
  On the other hand, from the NBS wave function for $|\br| < R$,
  we can define  a $k$-dependent local potential $U_k(\br)$
  and derive an associated  $k$-independent non-local potential
 $U(\br,\br^\prime)$ 
 with the use of the information of the  NBS wave functions for $E< E_{\rm th}$:
\begin{eqnarray}
  \left(\nabla_{\brs}^2 + k^2\right) \ \phi(\br;k)
  & \equiv & 2 \mu U_k(\br) \phi(\br;k) \label{eq:SE0}\\
  &=&
  2\mu
  \int d^3
   r^\prime
  \ U(\br,\br^\prime)
  \ \phi(\br^\prime;k), \label{eq:SE}
\end{eqnarray}
where $\mu=m_N/2$ denotes the reduced mass of the NN system.
 Derivation of Eq.(\ref{eq:SE}) from Eq.(\ref{eq:SE0}) 
 is given explicitly in Ref.~\citen{Aoki:2009ji}. Note also that
 an  equivalence theorem between $U_k(\br)$ and $U(\br,\br^\prime)$
 has been proved in a different manner in  Ref.~\citen{KR56}.
  In practical applications,   $U(\br,\br^\prime)$
   has an advantages over $U_k(\br)$; 
  its $k$-independence leads to the standard eigenvalue problem for the NBS wave function.
  Furthermore  its non-locality can be treated by the derivative expansion, 
$ U(\br,\br')=  V(\br, {\nabla}_{\brs})  \delta^3(\br-\br^\prime)$, with 
\begin{eqnarray} 
    V(\br,{\nabla}_{\brs})
    &=&
   \underbrace{
     V_{0}(r)
     + V_{\sigma}(r)
     (\bsigma_1 \cdot \bsigma_2)
     + V_{\rm T}(r) S_{12}
   }_{\rm LO}
   +
   \underbrace{
     V_{\rm LS}(r) \ \bL \cdot \bS
   }_{\rm NLO}
   + \mathcal{O}({\nabla^2}),
   \label{eq:derive}
\end{eqnarray}
  where $S_{12}  \equiv 3  (\bsigma_1 \cdot  \br)  (\bsigma_2
\cdot  \br  )/r^2-\bsigma_1  \cdot \bsigma_2$,   
$\bS \equiv \left(\bsigma_1 + \bsigma_2 \right)/2$ and
$\bL \equiv \br  \times (-i {\nabla}_{\brs})$
denote the tensor  operator, the total spin operator and the orbital angular momentum operator, respectively \cite{TW67,okubo}.

Since the total wave function has to be anti-symmetric under the exchange of two nucleons,
possible combinations of the total isospin $I$,  the total spin $S$ and
 the orbital angular momentum   $\ell$  are   restricted  to   four  cases,
 ($I, S, \ell$)=(1,0,even), (0,1,even), (1,1,odd) and (0,0,odd). Thus  
  we may omit the isospin $I$ indices in Eq.(\ref{eq:derive}).
 Note that the spin-singlet states ($S=0$) and spin-triplet states ($S=1$)
do not  mix with each  other, since the  isospin $I$ and  the parity
$P=(-1)^{\ell}$ are conserved for QCD with degenerate 2-flavors. 
To specify  two-nucleon  scattering states, we follow the standard
 notation, $^{2S+1}\ell_J$, with $J$ being the total angular momentum.    

\subsection{Spin-singlet potentials}
Let us first  consider the spin-singlet channel.  Since 
contributions from  $S_{12}$  and  $\bL \cdot \bS$ terms  are absent 
 in this case,  the Schr\"odinger equation reads
\begin{equation}
  \left(
  \nabla_{\brs}^2 + k^2
  \right)
  \phi(\br; k)
  =
  2\mu
  \left[
  V_{0}(r) - 3V_{\sigma}(r)
  + \left\{\nabla_{\brs}^2, V_{p^2}(r)\right\}
  + V_{L^2}(r) \bL^2
  + \mathcal{O}(\nabla^4) \right]
  \phi(\br; k).
  \label{eq:schrodinger.1s0}
\end{equation}
  Terms involving $2n$ derivatives such as $(\bL^2)^n$ and  $(\nabla^2)^{n}$ give N$^{2n}$LO
potentials.  (Note  that N$^{2n+1}$LO potentials are absent  in  the  spin-singlet channel.) 
At the  LO level,    the  Schr\"odinger
equation (\ref{eq:schrodinger.1s0})  reduces to
\begin{equation}
  \left(
  \nabla_{\brs}^2 + k^2
  \right)
  \phi(\br; k)
  =
  2\mu
  V_{\rm C,s}^{({\rm LO})}(r)
  \phi(\br; k),
  \label{eq:schrodinger.1s0.LO}
\end{equation}
with $V_{\rm C,s}^{({\rm LO})}(r)\equiv V_{0}(r) - 3V_{\sigma}(r)$.
Then the LO central potential in the spin-singlet channel is given by
\begin{equation}
V_{\rm C,s}^{({\rm LO})}(r)
  \equiv
  E  + \frac1{2\mu}
  \frac{\nabla_{\brs}^2 \phi(\br; k)}{\phi(\br; k)},
  \label{eq:central.potential.LO}
\end{equation}
where   $E \equiv \frac{k^2}{2\mu}$ is the effective kinetic energy between two nucleons.
 The above LO truncation works only when the right hand side of 
 Eq.(\ref{eq:central.potential.LO}) depends weakly on $k$ and $\ell$.
This will be checked explicitly in Sec.\ref{sec:numerical}
 at low energies and at low angular momentums through
  the comparisons, $(\ell=0, E \simeq 0$ MeV) vs. $(\ell=0, E \simeq 45$ MeV)
 and $(\ell=0, E \simeq 45$ MeV) vs. $(\ell=2, E \simeq 45$ MeV).

If $k$ and $\ell$ dependence in the spin-singlet channel becomes visible  as these values increase,
 it is a sign of the NNLO terms in Eq.(\ref{eq:schrodinger.1s0}). 
 Then the next step is to  determine NNLO 
   potentials $V_{\rm C,s}^{({\rm NNLO})}(r)$, $V_{p^2}^{({\rm NNLO})}(r)$
    and $V_{L^2}^{({\rm NNLO})}(r)$ through the NBS wave functions
    measured  with three different combinations of $k$ and $\ell$.
 Such a procedure continues to higher orders as $k$ and $\ell$ further increase. 
 A close analogy of this process is the renormalization-scale ($\kappa$) dependence
  in the perturvative series of quantum field theory;
 the artificial $\kappa$ dependence of scale-independent quantities
  is canceled order by order as we proceed to higher orders.

\subsection{Spin-triplet potentials}
For the spin-triplet channel,
the Schr\"odinger equation reads
\begin{equation}
  \left(
  \nabla_{\brs}^2 + k^2
  \right)
  \phi(\br; k)
  =
  2\mu
  \left[
     V_{0}(r)  +  V_{\sigma}(r)
    + V_{\rm T}(r) S_{12}
    + V_{\rm LS}(r) \bL \cdot\bS
    + \mathcal{O}(\nabla^2)
    \right]
  \phi(\br; k),
  \label{eq:STRI}
\end{equation}
 At the LO level, it reduces to
\begin{equation}
  \left(
  \nabla_{\brs}^2 + k^2
  \right)
  \phi(\br; k)
  =
  2\mu
  \left[
    V_{\rm C,t}^{({\rm LO})}(r)
    + V_{\rm T}^{({\rm LO})}(r) S_{12}
    \right]
  \phi(\br; k),
  \label{eq:schrodinger.3e.LO}
\end{equation}
where  $V_{\rm C,t}^{({\rm LO})}(r)\equiv V_{0}(r) +V_{\sigma}(r)$.

To  be  specific, we  restrict
ourselves to the case with $J^{P}=1^+$ NBS wave function 
 to which  two partial  waves contribute,  
i.e.,  $^3\!\,{\rm S}_1$ (S-wave [$\ell=0$]) and $^3\!\,{\rm D}_1$ (D-wave [$\ell=2$]).
As shown in Ref.~\citen{Aoki:2009ji},
\Eq{eq:schrodinger.3e.LO}    consists   of   two  independent equations 
\begin{equation}
  \left(
    \begin{array}{ll}
      {\cal P}\phi(\br; k) & {\cal P}S_{12}\phi(\br; k) \\
      {\cal Q}\phi(\br; k) & {\cal Q}S_{12}\phi(\br; k)
    \end{array}
    \right)
  \left(
  \begin{array}{c}
    V_{\rm C,t}^{({\rm LO})}(r) -k^2/2\mu \\
    V_{\rm T}^{({\rm LO})}(r)
  \end{array}
  \right)
  =
  \frac{\nabla_{\brs}^2}{2\mu}
  \left(
  \begin{array}{c}
    {\cal P}\phi(\br; k) \\
    {\cal Q}\phi(\br; k)
  \end{array}
  \right),
  \label{eq:coupled_equation.LO}
\end{equation}
where ${\cal P}$ (${\cal Q}$) is a projection to the $\ell = 0$ ($\ell = 2$) state.  
The LO potentials, $V_{\rm  C,t}^{({\rm LO})}(r)$   and  $V_{\rm  T}^{({\rm LO})}(r)$,   are  obtained  by solving this $2\times 2$ matrix equation  algebraically.

Spatial symmetry group of the hyper-cubic lattice is
 the cubic transformation group ${\rm SO}(3,{\mathbb{Z}})$ instead of the rotation group
  ${\rm SO}(3,{\mathbb{R}})$.  Here we employ the $J^P=T_1^+$ representation of 
  the ${\rm SO}(3,{\mathbb{Z}})$ for the wave function in the spin-triplet channel
  \footnote{Here $J$ is used to represent the quantum number of $ {\rm orbital}\otimes{\rm spin}$ even for the discrete group $SO(3,\mathbb{Z})$, and $P$ is the parity under the spatial reflection. }.
Since the spin-triplet belongs to the $T_1$ representation, the $J^P=T_1^+$ state in general contains orbital state $R$ which satisfies $T_1^+\in R\otimes T_1$.
Table~\ref{tab:product} in appendix~\ref{sec:cubic} gives $R=A_1^+$, $E^+$, $T_2^+$ and $T_1^+$. Among them we take the projection to the orbital $A_1^+$ representation for ${\cal P}$ as
\begin{eqnarray}
  {\cal P}\phi(\br; k)
  \equiv
  \frac1{24}
  \sum_{{\cal R} \in {\rm SO}(3,\mathbb{Z})}
  \phi({\cal R}^{-1}[\br]; k),
\end{eqnarray}
where the summation is  performed over  the cubic  group SO$(3,\mathbb{Z})$  with 24 elements.
The orbital $A_1$ representation is expected to be dominated by the S-wave  up  to
contamination of higher partial waves with $\ell \ge 4$.
  We employ ${\cal Q}=1-{\cal P}$ as a projection to non-$A_1^+$  orbital components
  composed of $E^+$, $T_2^+$ and $T_1^+$ representations.  
Non-$A_1^+$ orbital components are expected to be dominated by the D-wave
up to contamination of higher partial waves with $\ell \ge 4$.
Note  that $E^+$ and  $T_2^+$ contain  the $\ell = 2$  component, whereas
$T_1^+$ does not contain the $\ell = 2$ component.
 
If $k$ and $\ell$ dependence in the spin-triplet channel becomes visible  as these values increase,
 it is a sign of the NLO terms in Eq.(\ref{eq:STRI}).
Then the next step  is to  determine NLO 
   potentials  through the NBS wave functions
    measured  with several different combinations of $k$ and $\ell$.
 
\section{Finite-energy NN scattering on the lattice}
\label{sec:boundary}

To extract the NBS wave function on the lattice,
we start with the four-point nucleon correlation function,
\begin{eqnarray}
  G_{\alpha\beta}(\bx - \by,&t-t_0&; \mathcal{J}_{pn})
  \equiv
  \frac1{L_{\rm s}^3}
  \sum_{\br}
  \langle 0 |
  T[
    {p}_\alpha(\bx + \br,t)
    {n}_\beta(\by   + \br,t)
    \mathcal{J}(t_0)
  ]
  | 0 \rangle,
  \label{eq:four-point-correlator} \\
  &\simeq&
  \phi_{\alpha\beta}(\bx-\by; k) \langle B=2;k| \mathcal{J}(0)|0 \rangle
  e^{-W (t-t_0)}, \  \ t-t_0 \gg 1,
  \label{eq:4-point} 
\end{eqnarray}
where  the summation over  $\br$  is performed  to select the two nucleon system
with total spatial momentum zero, 
$\mathcal{J}(t_0)$  denotes  a two-nucleon  source  located at  $t=t_0$,
whose explicit  form will be  specified below. 
 The 
 relativistic energy and associated asymptotic momentum of the 
  ``ground" state of the $B=2$ system are denoted by 
  $W$ and $k$, respectively. As for the sink operators, $p(x)$ and $n(x)$,
  we employ the following local composite operators,
\begin{equation}
  p(x)
  \equiv
  \epsilon_{abc}
  \left(
  u_a^T(x)
  C\gamma_5
  d_b(x)
  \right)
  u_c(x),
  \ \ \ \ 
  n(x)
  \equiv
  \epsilon_{abc}
  \left(
  u_a^T(x)
  C\gamma_5
  d_b(x)
  \right)
  d_c(x),
\end{equation}
where $a,b,c$ are  color indices. 

The NBS  wave function at  $E \simeq  0$ MeV is  generated under
 the periodic  boundary condition (PBC), which is  imposed on the
quark operators  along the spatial  directions.  With the PBC, the momentum  of a
single nucleon is discretized as  $k_i = 2\pi n_i/L_{\rm s}$ with
 $n_i \in \mathbb{Z}$. 
Hence, the  lowest lying  state of  the two nucleon  system in  the CM
frame roughly corresponds to the  state where two nucleons are weakly
interacting  with  relative  momentum  of  $k_i \simeq  0$  MeV.   The
effective kinetic energy of such a state is $E \equiv k^2/m_N
\simeq 0 $ MeV.
The NBS  wave function at $E \simeq 45$ MeV  is generated under
 the anti-periodic boundary condition (APBC).
Since the  nucleon also obeys the APBC,  the spatial momentum  of a single
nucleon is discretized  as $k_i = (2 n_i +  1)\pi/L_{\rm s}$ with 
$n_i \in \mathbb{Z}$.
Hence, the  lowest lying  state of  the two nucleon  system in  the CM
frame roughly corresponds to the  state where two nucleons are weakly
interacting with relative momentum  of $k_i \simeq \pm \pi/L_{\rm s}$.
For the lowest lying state with  $L_{\rm s}\simeq 4.4$ fm, 
the spatial momentum of a nucleon amounts to $|\bk | \simeq
\sqrt{3}  \pi/L_{\rm s}  \simeq  245$ MeV,  which  corresponds to 
 $E \simeq 45$ MeV in  our setup with
$m_N \simeq 1.33$ GeV.

As for the source operators of the two nucleon system, we employ
\begin{equation}
  \mathcal{J}_{\alpha\beta}(f)
  \equiv
  \bar{P}_{\alpha}(f)
  \bar{N}_{\beta}(f),
\end{equation}
where $\bar{P}_{\alpha}(f)$  and  $\bar{N}_{\beta}(f)$  associated with  a
source function $f(\bx)$ are given as
\begin{eqnarray}
  \bar{P}(f)
  &\equiv&
  \epsilon_{abc}
  \left(
  \bar{U}_a(f)
  C\gamma_5
  \bar{D}_b^T(f)
  \right)
  \bar{U}_c(f),
  \nonumber
  \\
  \bar{N}(f)
  &\equiv&
  \epsilon_{abc}
  \left(
  \bar{U}_a(f)
  C\gamma_5
  \bar{D}_b^T(f)
  \right)
  \bar{D}_c(f).
\end{eqnarray} 
 Here  the source operators for $u$ and $d$ quarks are given by
 \begin{equation}
  \bar{U}(f)
  \equiv
  \sum_{\bxs}
  \bar{u}(\bx)
  f(\bx),
  \hspace{2em}
  \bar{D}(f)
  \equiv
  \sum_{\bxs}
  \bar{d}(\bx)
  f(\bx).
\end{equation}

An element ${\cal R}$ of the cubic group SO(3,$\mathbb{Z}$) rotates the quark field operator as
\begin{equation}
  \bar{q}(\bx)\mapsto \bar{q}({\cal R}^{-1}\bx) \Lambda({\cal R}^{-1}),
  \label{eq.1S0quark}
\end{equation}
where  $\Lambda$   denotes  the 4-component  spinor  representation   of  $O$  as
$\Lambda(e^{\omega}) \equiv \exp \left(- \frac{i}{4} \sigma_{ij} \omega^{ij}
\right)$ with $\sigma_{\mu\nu} \equiv \frac{i}{2}\left[\gamma_{\mu},   \gamma_{\nu}\right]$.
 This
leads to the transformation property of $\mathcal{J}_{\alpha\beta}(f)$ as
\begin{equation}
  \mathcal{J}_{\alpha\beta}(f)
  \mapsto
  \mathcal{J}_{\alpha'\beta'}({\cal R}^{-1}\circ f)
  \Lambda_{\alpha'\alpha}({\cal R}^{-1})
  \Lambda_{\beta' \beta} ({\cal R}^{-1}),
\end{equation}
where $({\cal R}^{-1}\circ f)(\bx) \equiv f({\cal R} \bx)$.

To consider $J=0$ and 1,
 it is convenient to introduce a source operator
 which has definite $J$ and  $M$ with $J=0+S$ and $M=0+S_z$
  to  construct NBS wave functions in the $^1\!\,{\rm S}_0$ 
  and $^3\!\,{\rm S}_1-^3\!\,{\rm D}_1$  channels:
\begin{equation}
  \mathcal{J}^{(J,M)}(f)
  \equiv
  \frac1{24}
  \sum_{{\cal R}\in {\rm SO}(3,\mathbb{Z})}
  \mathcal{J}_{\alpha\beta}({\cal R}^{-1}\circ f)
  \cdot
  P^{(S=J,S_z=M)}_{\alpha\beta},
  \label{eq.l=0.source}
\end{equation}
where $P_{\alpha\beta}^{(S,S_z)}$ denotes the spin projection operator
defined         as        $P_{\alpha\beta}^{(S=0,S_z=0)}        \equiv
(\sigma_2)_{\alpha\beta}/\sqrt{2}$,      $P_{\alpha\beta}^{(S=1,S_z=M)}
\equiv (\sigma_2\sigma_{M})_{\alpha\beta}/\sqrt{2}$ with $M=\pm 1, 0$, where
 we take only the upper components of the Dirac indices for simplicity.

  For  the PBC, we  employ a  flat
wall (f-wall) source,
\begin{equation}
  f^{({\rm f-wall})}(\br) = 1,
\end{equation}
which is invariant under the rotation ${\cal R}$. Then,
  \Eq{eq.l=0.source} reduces to
\begin{equation}
  \mathcal{J}^{(J,M)}(f^{({\rm f-wall})})
  =
 \bar{P}_{\alpha}(f^{({\rm f-wall})})
  \bar{N}_{\beta} (f^{({\rm f-wall})}) \cdot  P^{(S=J,S_z=M)}_{\alpha\beta},
\end{equation}
which couples dominantly to the ground state ($\bk=(0,0,0) \pi/L_{\rm s}$) in the PBC.

For the APBC, we utilize a set of momentum wall sources 
$f^{\rm (m-wall)} = \{f^{(i)}\}_{i=0-3}$ with
\begin{eqnarray}
  f^{(0)}(\br) &\equiv& \cos((+x+y+z)\pi/L_{\rm s}), \nonumber\\
  f^{(1)}(\br) &\equiv& \cos((-x+y+z)\pi/L_{\rm s}), \nonumber\\
  f^{(2)}(\br) &\equiv& \cos((-x-y+z)\pi/L_{\rm s}), \nonumber\\
  f^{(3)}(\br) &\equiv& \cos((+x-y+z)\pi/L_{\rm s}),
  \label{eq:source_APBC}
\end{eqnarray}
where  the cosine function is chosen to create positive parity states.

The cubic  group acts on  these functions  as permutation, 
 which is characterized  by  the  cubic   group  representation,
$A_1^+\oplus T_2^+$.
By taking the $A_1^+$ part,   \Eq{eq.l=0.source} becomes
\begin{equation}
  \mathcal{J}^{(J,M)}(f^{({\rm m-wall})};A_1^+)
  \equiv
  \frac1{4}
  \sum_{j=0}^3
  \bar{P}_{\alpha}(f^{(j)})
  \bar{N}_{\beta }(f^{(j)})
  \cdot
    P^{(S=J,S_z=M)}_{\alpha\beta},
 \label{eq:J-A1plus}   
\end{equation}
which couples dominantly to the ground state ($\bk=(1,1,1) \pi/L_{\rm s}$) in  the APBC.
Since this source operator is not translational invariant, 
it is practically important to perform a summation over $\br$ at the sink side in
\Eq{eq:four-point-correlator} to pick up zero spatial momentum states.
Instead of  \Eq{eq:source_APBC}, 
one may choose  a simpler  cosine-type  function
 \begin{equation}
  f(\br)
  \equiv
  \cos(\pi x/L_{\rm s})
  \cos(\pi y/L_{\rm s})
  \cos(\pi z/L_{\rm s}),
  \label{eq:source_APBC2}
\end{equation}
which gives a source operator coupled to the ground state 
($\bk=(1,1,1) \pi/L_{\rm s}$) in the APBC. 
 However, it receives a contamination from the 
 coupling with the  first excited state ($\bk  = (3,1,1)\pi/L_{\rm s}$).
 In contrast, the source
operator  with \Eq{eq:source_APBC}  has  an overlap  neither with  the
first  excited state  nor  the  second excited  state, and receives
 contamination only from  the  third
excited state ($\bk = (3,3,3)\pi/L_{\rm s}$). 
Therefore,  signal for the ground state is better for
 \Eq{eq:source_APBC} than that for  \Eq{eq:source_APBC2}.

Since \Eq{eq:source_APBC} contains  $T_2^+$ component, 
it can be also used to generate the state in 
 the $^1\!\,{\rm D}_2$ channel, which is employed to study  the  $\ell$
 dependence  of $V^{({\rm LO})}_{\rm C,s}(r)$.
A general projection formula for the source operator in the spin-singlet
sector given in \Eq{eq:projection.s=0.general} leads to
\begin{equation}
  \mathcal{J}^{(J=2,M)}(f^{({\rm m-wall})};T_2^+)
  =
  \frac1{4}
\sum_{j=0}^{3}
  e^{i Mj \pi/2}
  \bar{P}_{\alpha}(f^{(j)})
  \bar{N}_{\beta} (f^{(j)}) \cdot  
   P^{(S=0,S_z=0)}_{\alpha\beta} 
  \label{eq:source_1d2}
\end{equation}
for the $T_2^+$ representation,
where $M$ takes 2 and $\pm 1$ (modulo  4). See Appendix
\ref{sec:D-wave-op} and \ref{sec:c4} for more details.  
In the  actual numerical calculation,  we take linear  combinations of
\Eq{eq:source_1d2} to make them into real basis as
\begin{eqnarray}
  \mathcal{J}^{J=2, xy}
  &\equiv&
  \mathcal{J}^{J=2, M=2}
  \nonumber
  \\
  \mathcal{J}^{J=2, yz}
  &\equiv&
  \frac{i}{\sqrt{2}}
  \left(
  \mathcal{J}^{J=2, M=-1}
  +
  \mathcal{J}^{J=2, M=1}
  \right)
  \nonumber
  \\
  \mathcal{J}^{J=2, zx}
  &\equiv&
  \frac1{\sqrt{2}}
  \left(
  \mathcal{J}^{J=2, M=-1}
  -
  \mathcal{J}^{J=2, M=1}
  \right).
\end{eqnarray}

\section{Numerical results}
\label{sec:numerical}

\subsection{Lattice setup}
Employing  the standard plaquette gauge  action on a  $32^3 \times 48$
lattice at $\beta=5.7$, quenched gauge configurations  are generated
by the heat-bath algorithm with the over-relaxation.
We accumulate 4000 configurations separated  by   200 sweeps. 
The standard Wilson quark action  is used to calculate quark 
propagators with the hopping parameter $\kappa=0.1665$.
 The Dirichlet boundary  condition in the temporal direction is imposed
  at $t-t_0 = \pm 24$. 
  The  nucleon four-point  correlation functions  are measured  for  both $t-t_0>0$  and
$t-t_0<0$  to  improve  the statistics  by using  the
time-reversal and charge conjugation symmetries\cite{Aoki:2009ji}. 
 Either PBC or APBC is taken in the spatial direction:  In the former case,
we use four sources at $t_0=0, 8, 16, 24$  to improve the statistics.
 These calculations are performed on Blue Gene/L at KEK.

From the  rho meson mass in  the chiral limit, the  lattice spacing 
is determined to be $a^{-1}=1.44(2)$ GeV  ($a \simeq 0.137$ fm), which leads to 
 $L_{\rm  s}=32a \simeq 4.4$ fm. 
  Our $\kappa$ corresponds to the pion mass $m_\pi\simeq  0.53$ GeV  
  and the  nucleon mass  $m_N \simeq  1.33$ GeV\cite{Fukugita:1994ve}.   
 After examining  the stability of the $NN$ potentials against the variation of 
 $t-t_0$, we chose 
  the wave functions and potentials  at $t-t_0=9$ in all the plots shown in this paper.
\subsection{The NBS wave functions}
\begin{figure}[tb]
\begin{center}
\includegraphics[width=6.5cm]{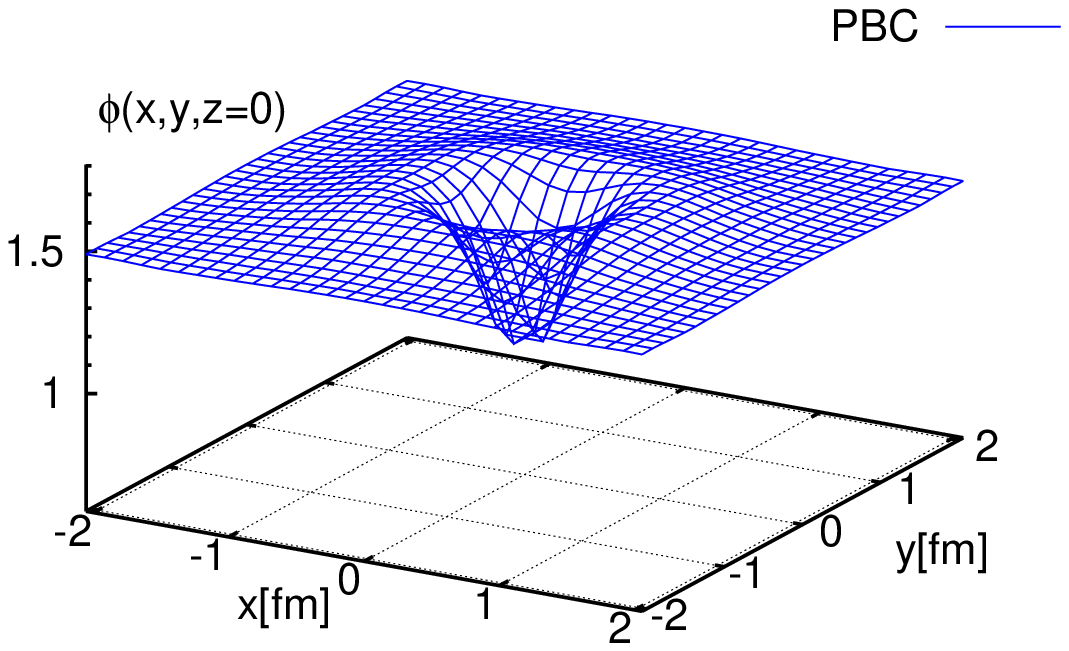}
\includegraphics[width=6.5cm]{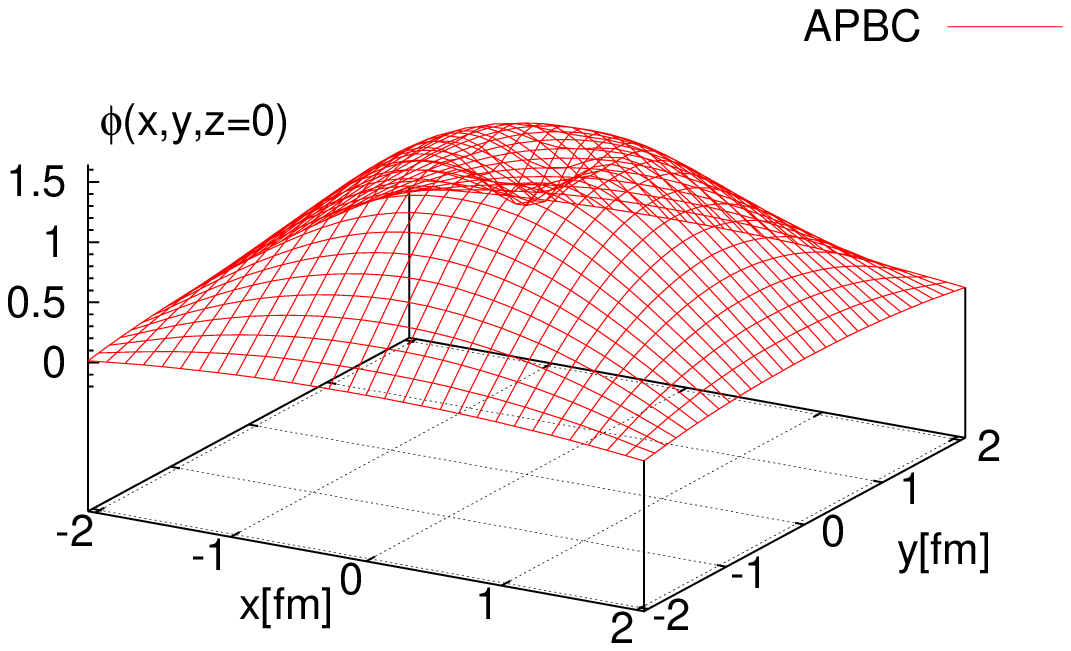}
\end{center}
  \caption{(Left) The  NBS wave function for the spin-singlet and the orbital $A_1^+$ channel at $E\simeq 0$  MeV with the PBC. (Right)  The NBS wave  function in the same channel but at $E\simeq 45$ MeV with the APBC. Both wave functions are normalized as $\phi(r=0)=1$. }
  \label{fig:1S0wave_3d}
\end{figure}

\begin{figure}[tb]
\begin{center}
\includegraphics[width=6.5cm]{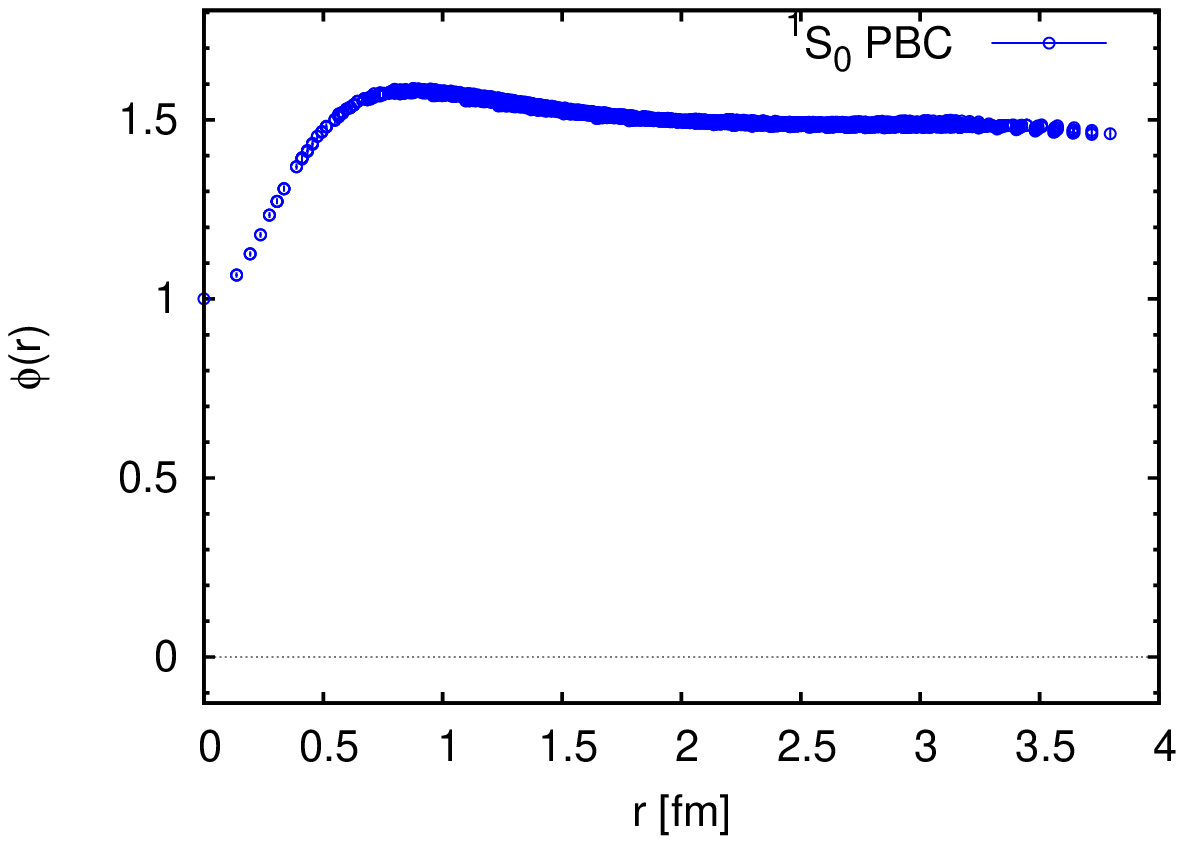}
\includegraphics[width=6.5cm]{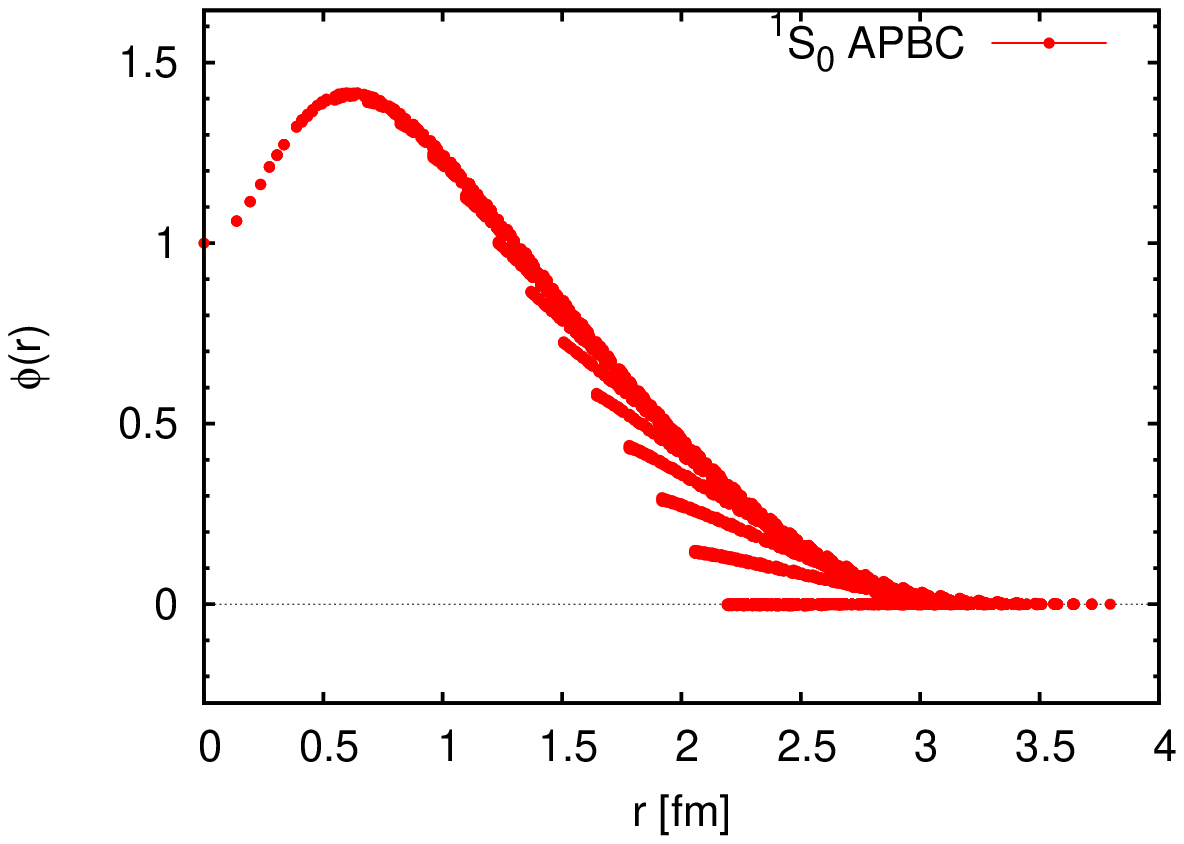}
\end{center}  
  \caption{Same NBS wave functions as in 
   \Fig{fig:1S0wave_3d} but as a function of $r$. }
  \label{fig:1S0wave}
\end{figure}

\Fig{fig:1S0wave_3d}(Left) and  (Right) show three dimensional plots of the  NBS wave functions
 $\phi(x,y,z=0)$
 for the spin-singlet and the orbital $A_1^+$ channel ($\simeq ^1\!{\rm S}_0$ channel)
  at $E\simeq 0$  MeV
  and at $E\simeq 45$  MeV,  respectively.   We observe that  they behave rather
   differently:
  The wave function for the PBC is almost constant  at long distances,
  which indicates that the asymptotic momentum is nearly zero. On the other hand, 
the wave function for the APBC decreases continuously to zero at long distances,
 since the
  wave function in orbital $A_1^+$ state must vanish on the boundary in APBC.
This can be seen, for example,  by using a $\pi$ rotation around the  x-axis followed by  the spatial reflection as
$
  \phi(x,y,z)
  =
  \phi(x,-y,-z)
  =
  \phi(-x,y,z)
  =
  -\phi(L_{\rm s}-x, y,z),
$
which leads to
$
  \phi(L_{\rm s}/2, y,z)
  =
  -\phi(L_{\rm s}/2, y,z)
  =
  0.
$

\begin{figure}[bt]
\begin{center}
\includegraphics[width=6.5cm]{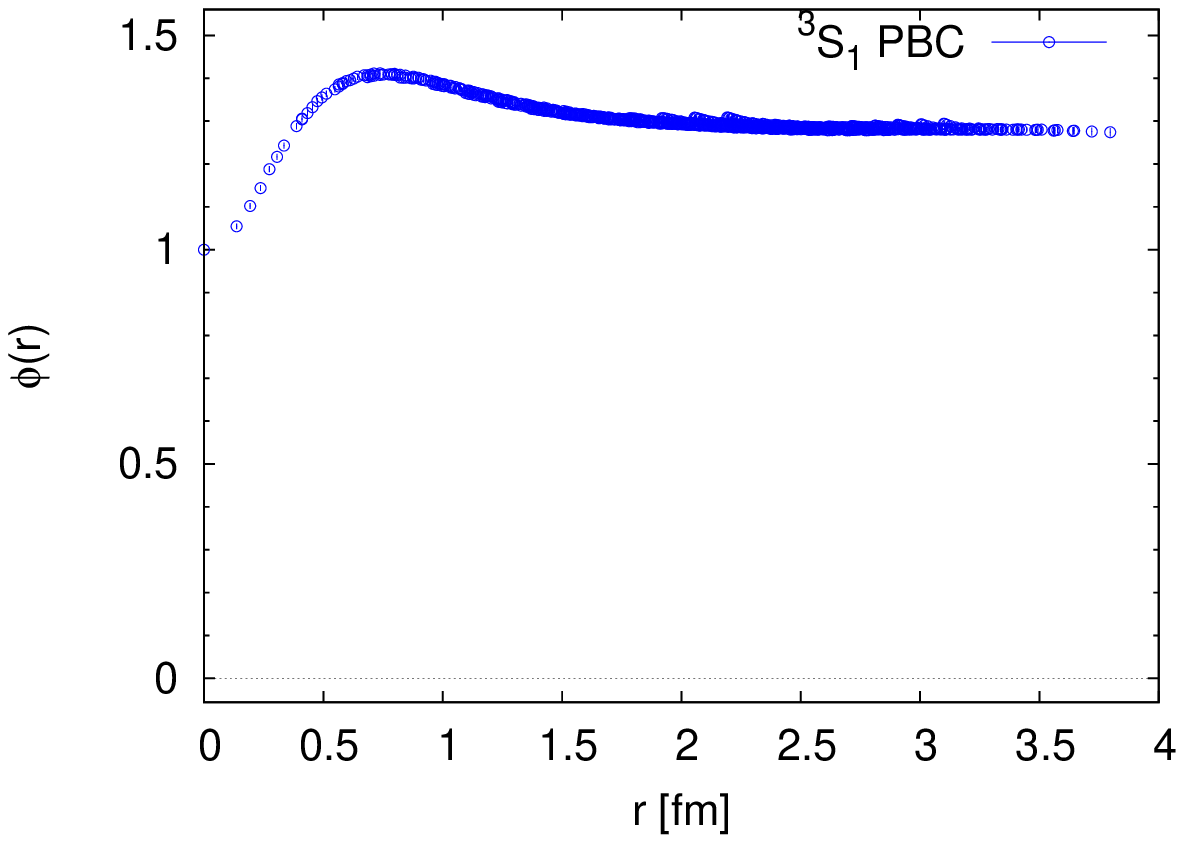}
\includegraphics[width=6.5cm]{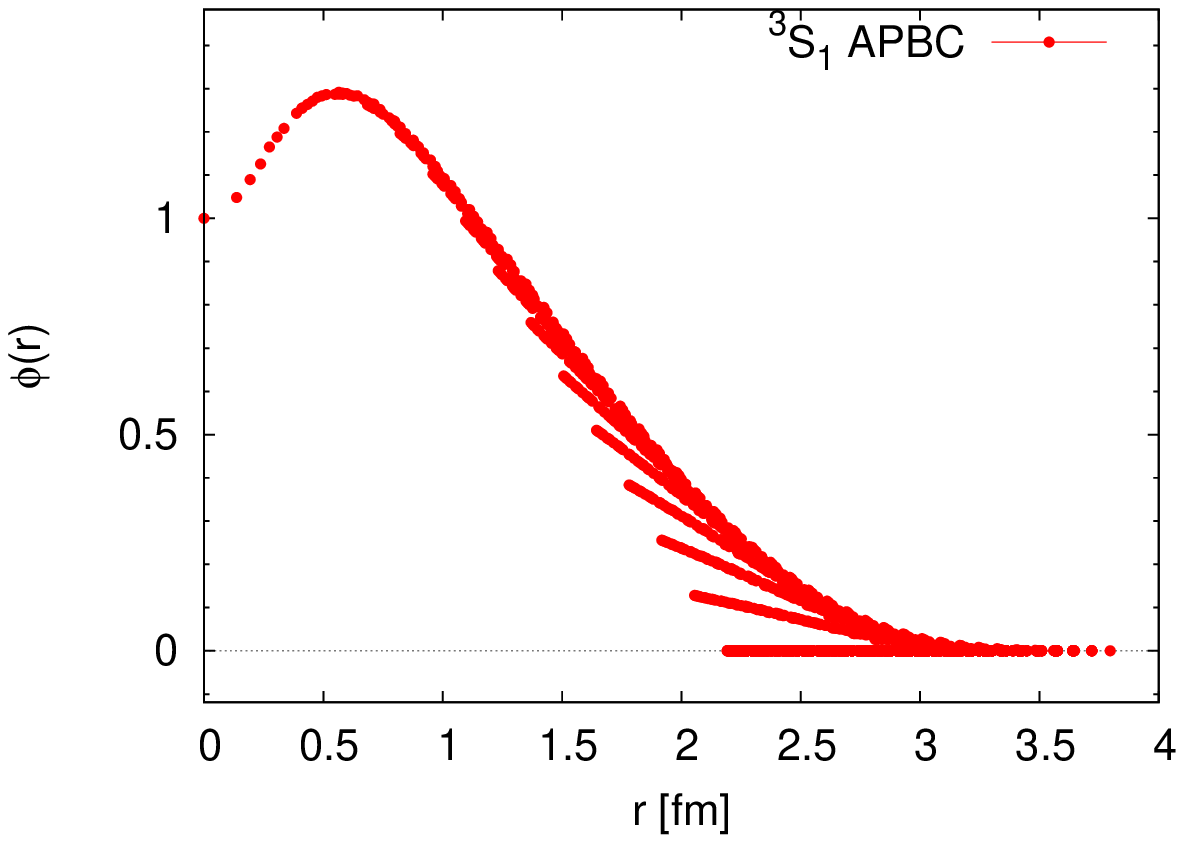}
\end{center}
  \caption{(Left) The  NBS wave function  in the spin-triplet and the orbital $A_1^+$ channel  at $E\simeq 0$  MeV with the PBC. (Right)  The same at  $E\simeq 45$ MeV with the APBC.
  Both wave functions are normalized as $\phi(r=0)=1$. }
  \label{fig:3S1wave}
\end{figure}

 In \Fig{fig:1S0wave}, the same wave functions as 
  \Fig{fig:1S0wave_3d} are plotted as a function of $r$.
  Violation of rotational symmetry due to the square lattice can be seen
   explicitly through the multi-valuedness of the wave function  
   at large  $r$ for the  APBC.
 Shown in \Fig{fig:3S1wave} are the similar comparison of  NBS  wave  functions
 between the PBC and the APBC  in 
  the spin-triplet   and the orbital $A_1^+$ channel ($\simeq ^3\!{\rm S}_1$ channel).

\begin{figure}[tb]
\begin{center}
\includegraphics[width=6.5cm]{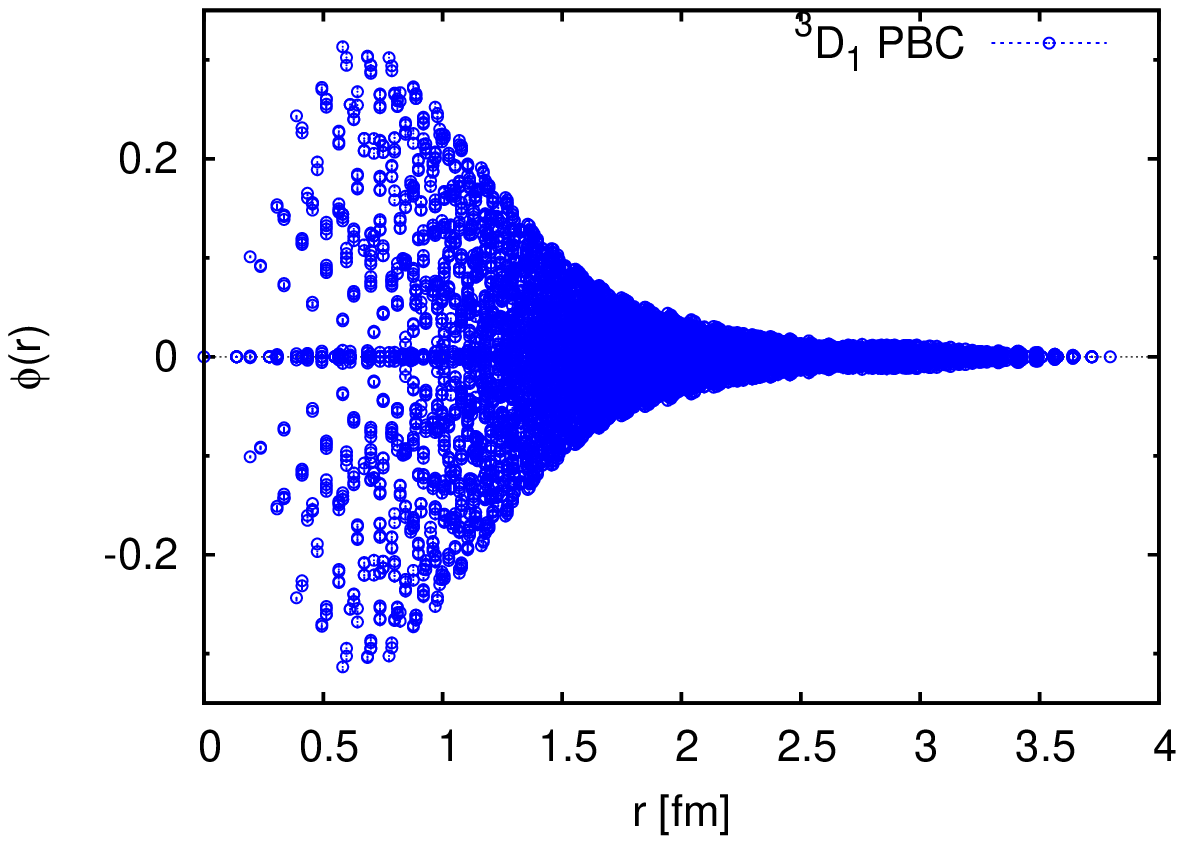}
\includegraphics[width=6.5cm]{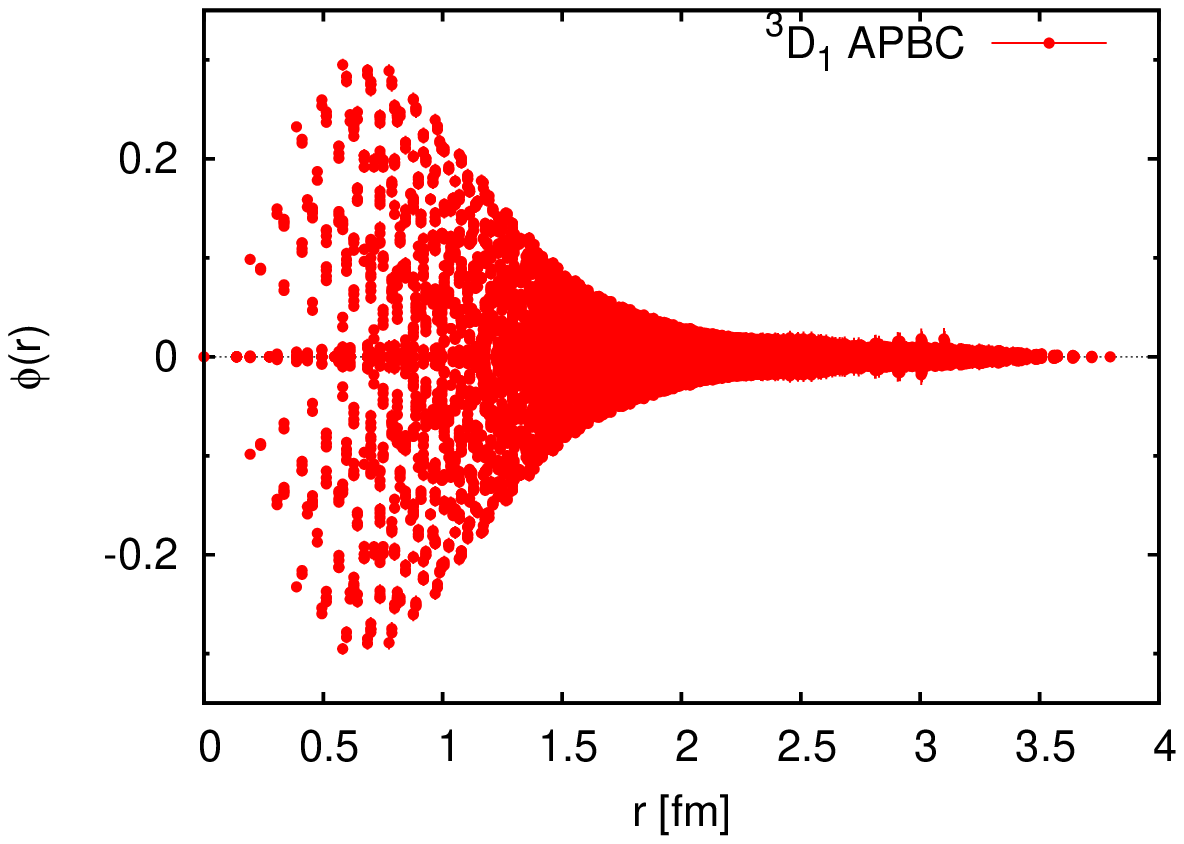}
\includegraphics[width=6.5cm]{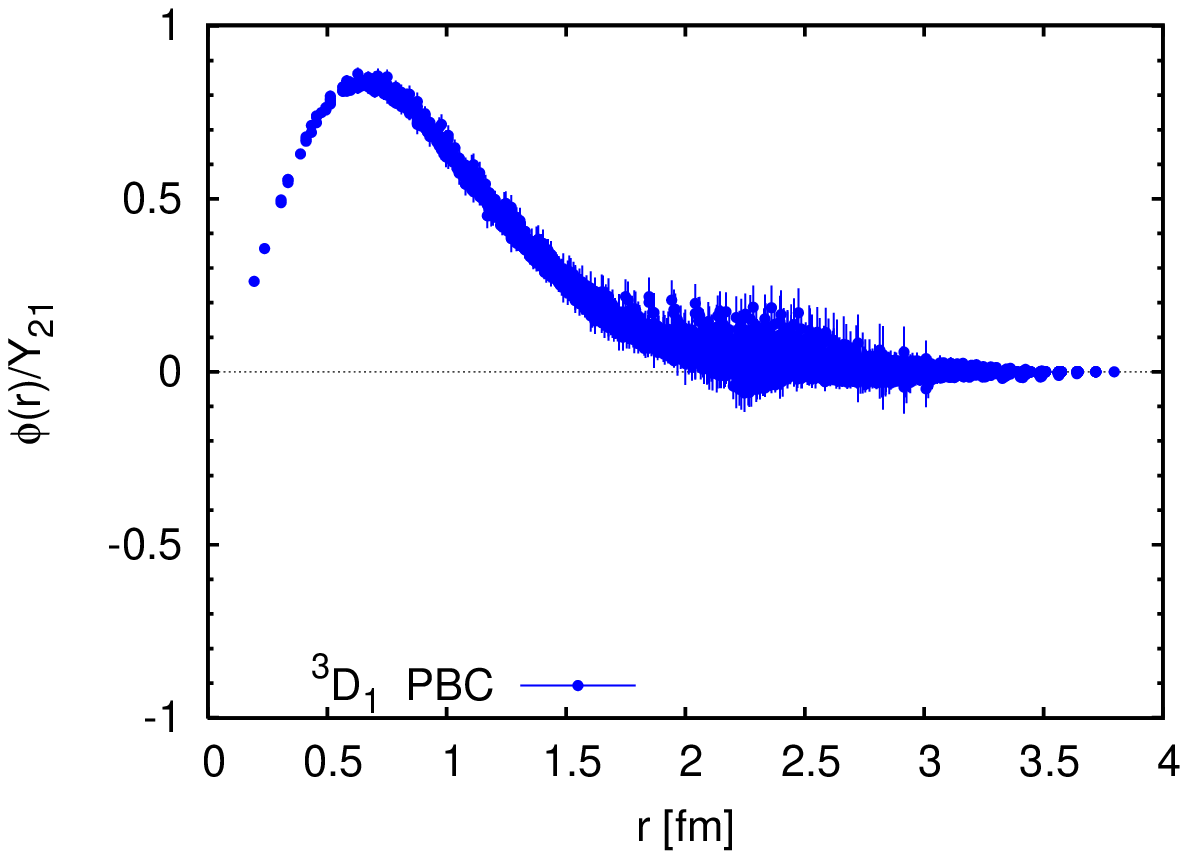}
\includegraphics[width=6.5cm]{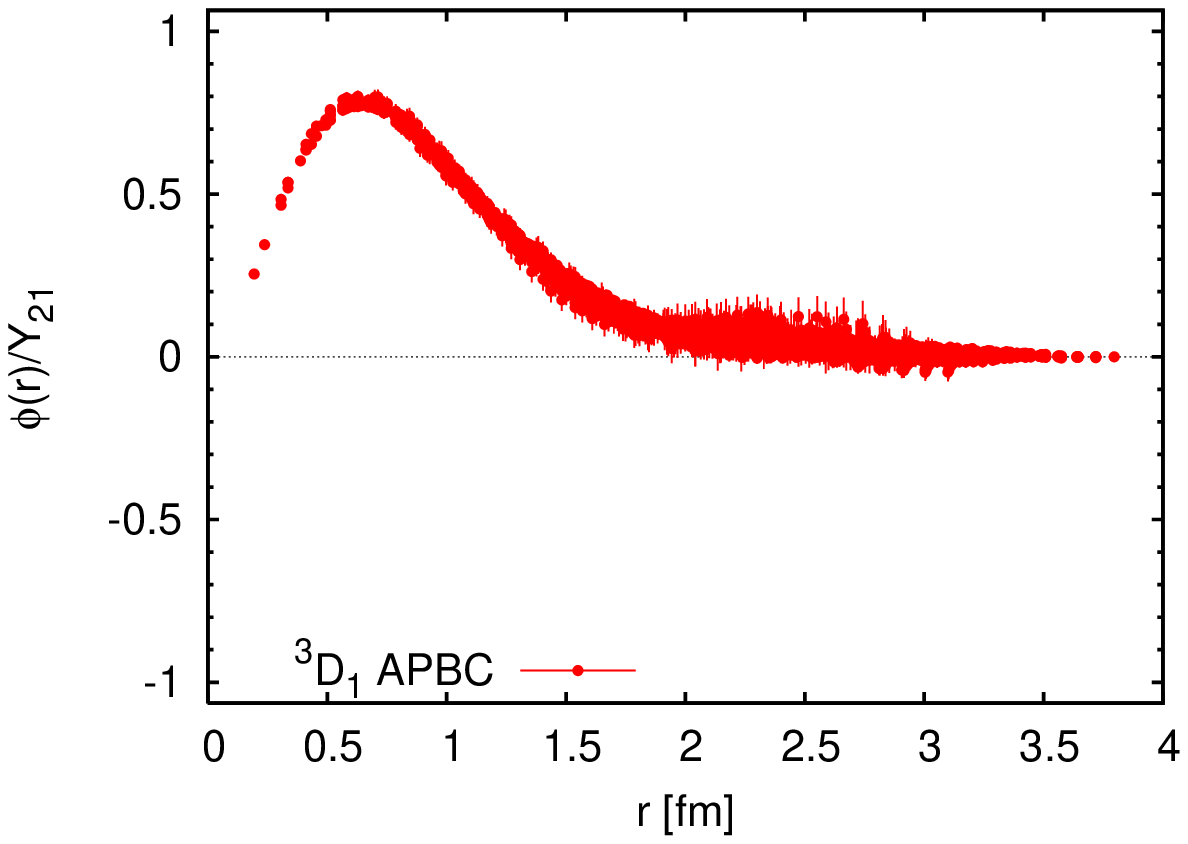}
\end{center}
  \caption{(Upper-Left) The NBS wave function ${\rm Re} \phi_{\downarrow\downarrow}(\br)$ for the spin-triplet and the
 orbital $T_2^+$ channel at $E \simeq 0$  MeV with the PBC.
    (Upper-Right) The same NBS wave function but at $E\simeq 45$  MeV with the APBC.
    (Lower-Left) 
       The NBS wave function $\phi_{\downarrow\downarrow}$ divided by the   spherical   harmonics   $Y_{21}(\theta,\phi)$.
(Lower-Right) Same as the left figure but at $E \simeq 45$ MeV with the
APBC. Normalization of these wave functions is fixed uniquely once the
  normalization of the S-wave part is fixed as given in
  Fig.\ref{fig:1S0wave} and Fig.\ref{fig:3S1wave}.  
}
  \label{fig:3D1wave}
\end{figure}

 In \Fig{fig:3D1wave}(Upper),  
 we plot the NBS wave functions for the
 spin-triplet and the orbital $T_2^+$ channel ($\simeq ^3\!{\rm D}_1$ channel).
   They are highly multi-valued  as  functions  of  $r$ 
   at all distances simply  due to the  angular dependence of the orbital $T_2^+$ 
  representation. To extract the radial part only,  
  we divide the wave functions by $Y_{2,m}(\theta,\phi)$
   assuming that the angular dependence is dominated by the 
  $\ell=2$ component. The results are shown in \Fig{fig:3D1wave}(Lower):
  almost single-valued radial wave functions are obtained for both PBC and APBC
   cases.  

\subsection{LO potentials for different energies}

\begin{figure}[tb]
\begin{center}
\includegraphics[width=6.5cm]{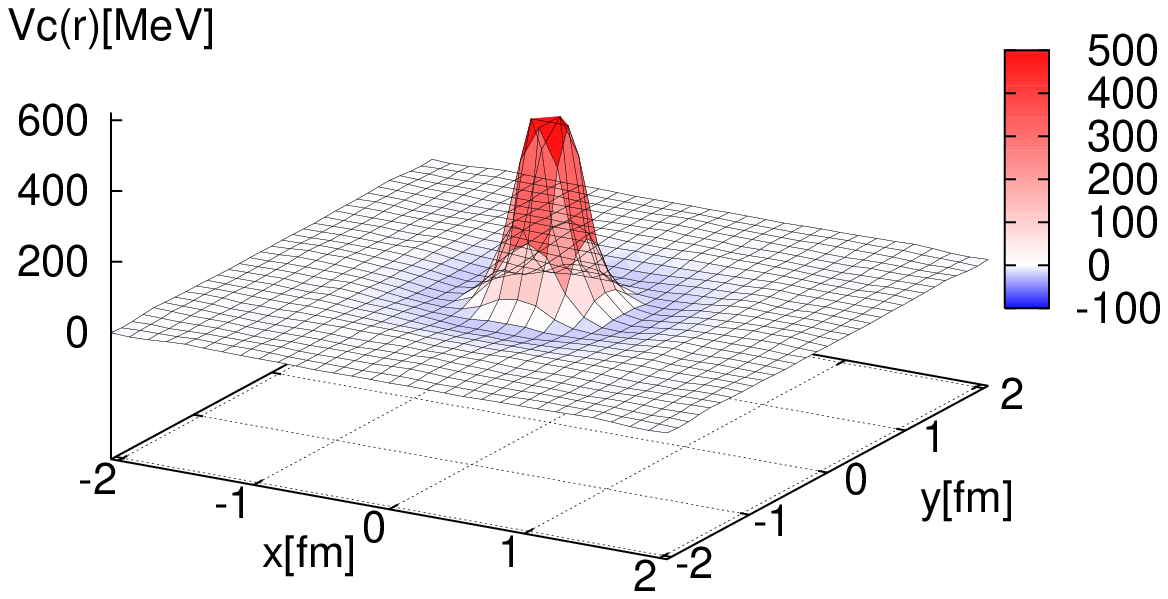}
\includegraphics[width=6.5cm]{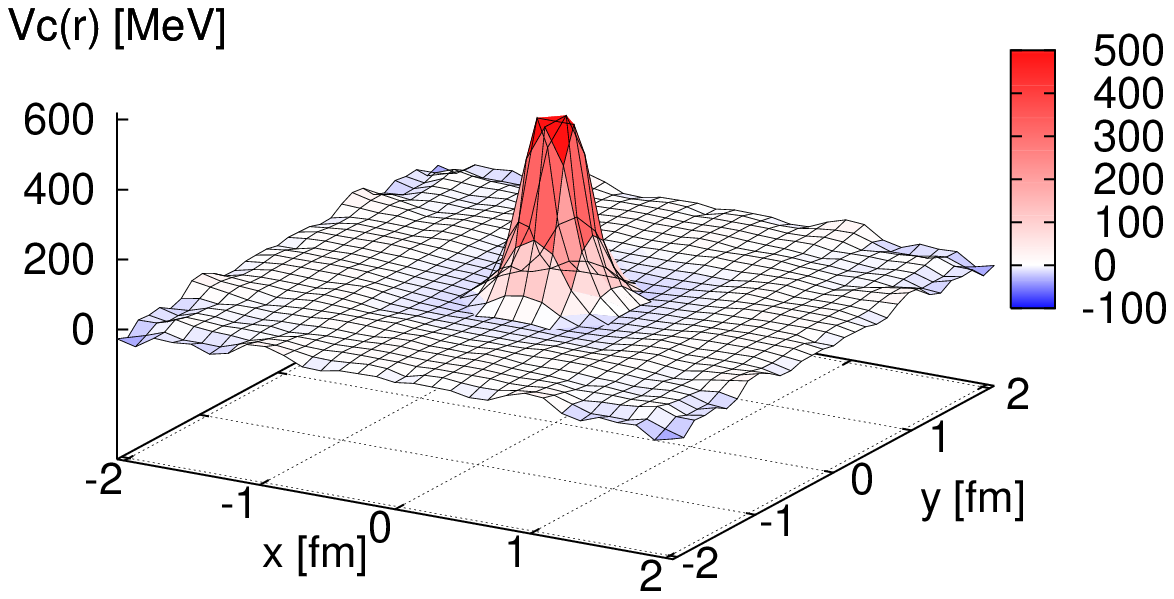}
\end{center}
  \caption{ The spin-singlet central potential, $V_{\rm C,s}^{\rm LO}(r)$, (Left) at $E\simeq 0$ MeV and (Right) at $E\simeq 45$ MeV . }
  \label{fig:singlet_3d}
\end{figure}

The leading order potentials are extracted from the NBS wave functions according to  
\Eq{eq:central.potential.LO} and \Eq{eq:coupled_equation.LO} for  the   spin-singlet  
 and spin-triplet channels, respectively.
In order to obtain LO potentials, we need to determine the value of $E=k^2/(2\mu)$ either from 
the large $t$ behavior of the NN correlation function or the large $r$ behavior 
of the NBS wave function.\footnote{We note here that a new method to obtain the potentials 
 by using the $t$-dependent Schr\"{o}dinger equation has been also proposed recently \cite{t-dep}.}
 It turns out 
that the values of $E$ from both determinations
are roughly  equal to their free values, i.e., $E  \simeq 0$
MeV for the PBC  and $E \simeq 45$ MeV  for the APBC, within statistical and 
 systematic errors.
Therefore, we adopt  these  free  values  as characteristic $E$ 
 in  extracting the  LO central potentials.
 Note that the tensor potential is free from the uncertainty of $E$ as can be seen
  from   Eq.(\ref{eq:coupled_equation.LO}).

\begin{figure}[tb]
\begin{center} 
\includegraphics[width=0.45\textwidth]{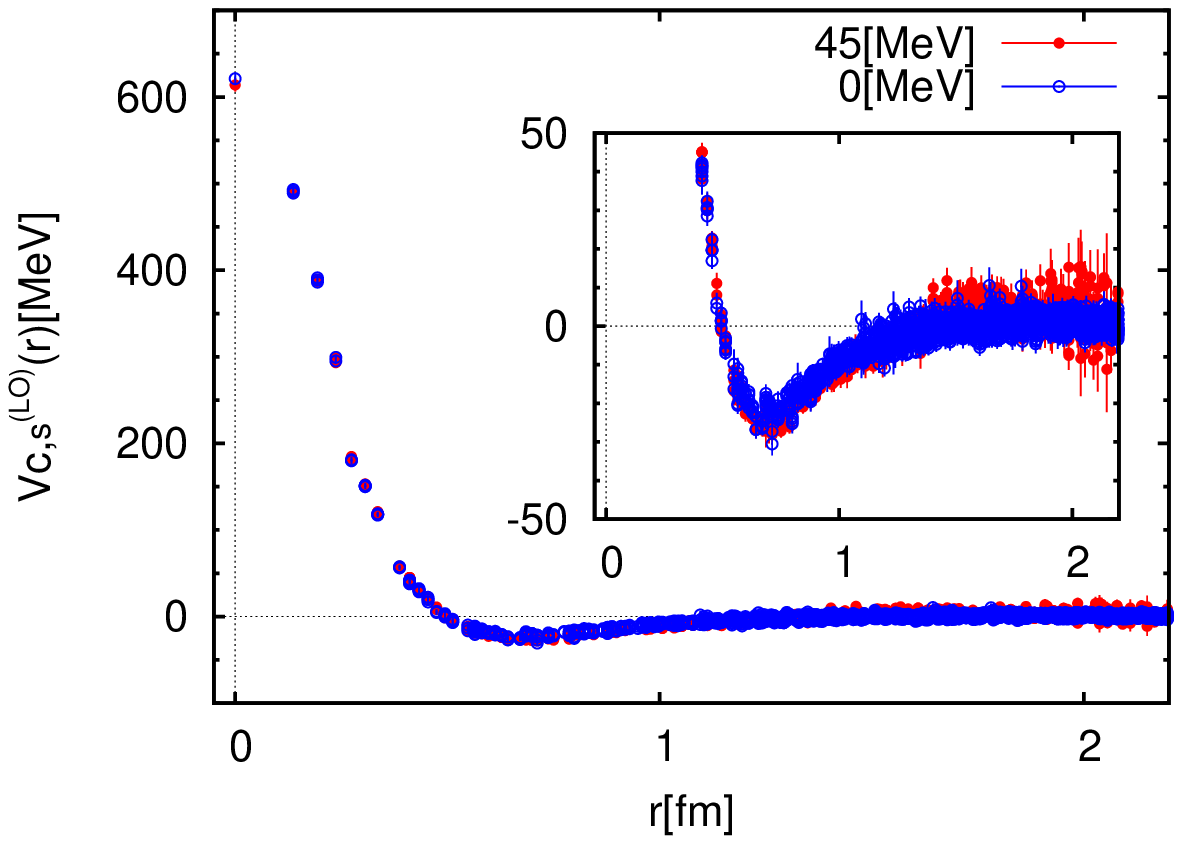}
\includegraphics[width=0.45\textwidth]{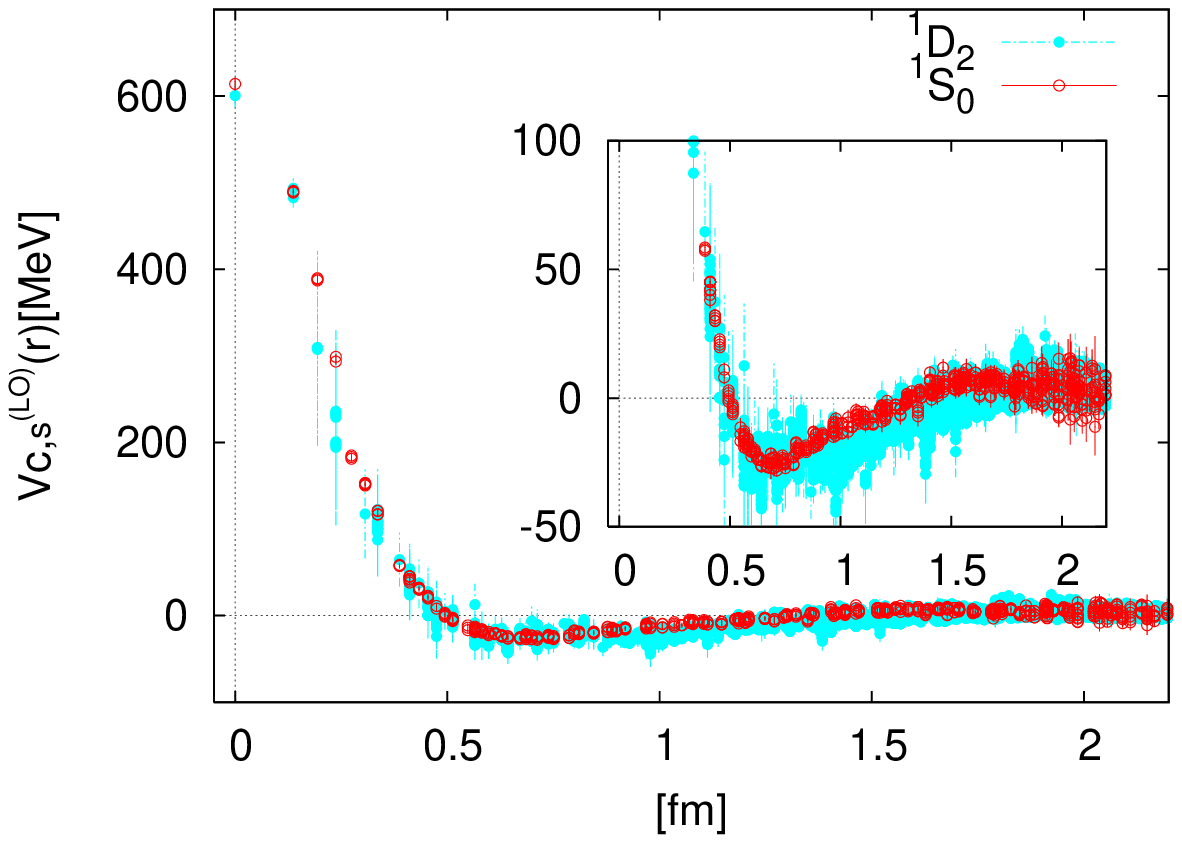}
\end{center}       
  \caption{ (Left) The  LO central potential $V_{\rm  C,s}^{\rm (LO)}(r)$ for the spin-singlet and the orbital $A_1^+$ channel as a function of $r$ at $E\simeq 0$ 45 MeV (red solid circles) and at $E\simeq 0$ MeV (blue open circles).
(Right)  The  LO central potential $V_{\rm  C,s}^{\rm (LO)}(r)$ for the spin-singlet  channel as a function of $r$ at $E\simeq$ 45 MeV , determined from the orbital $A_1^+$ representation
  (red open circles) and from the $T_2^+$ representation (cray solid circles). 
  }
  \label{fig:singlet_APBCvsPBC}
\end{figure}

\begin{figure}[tb]
\begin{center} 
\includegraphics[width=6.5cm]{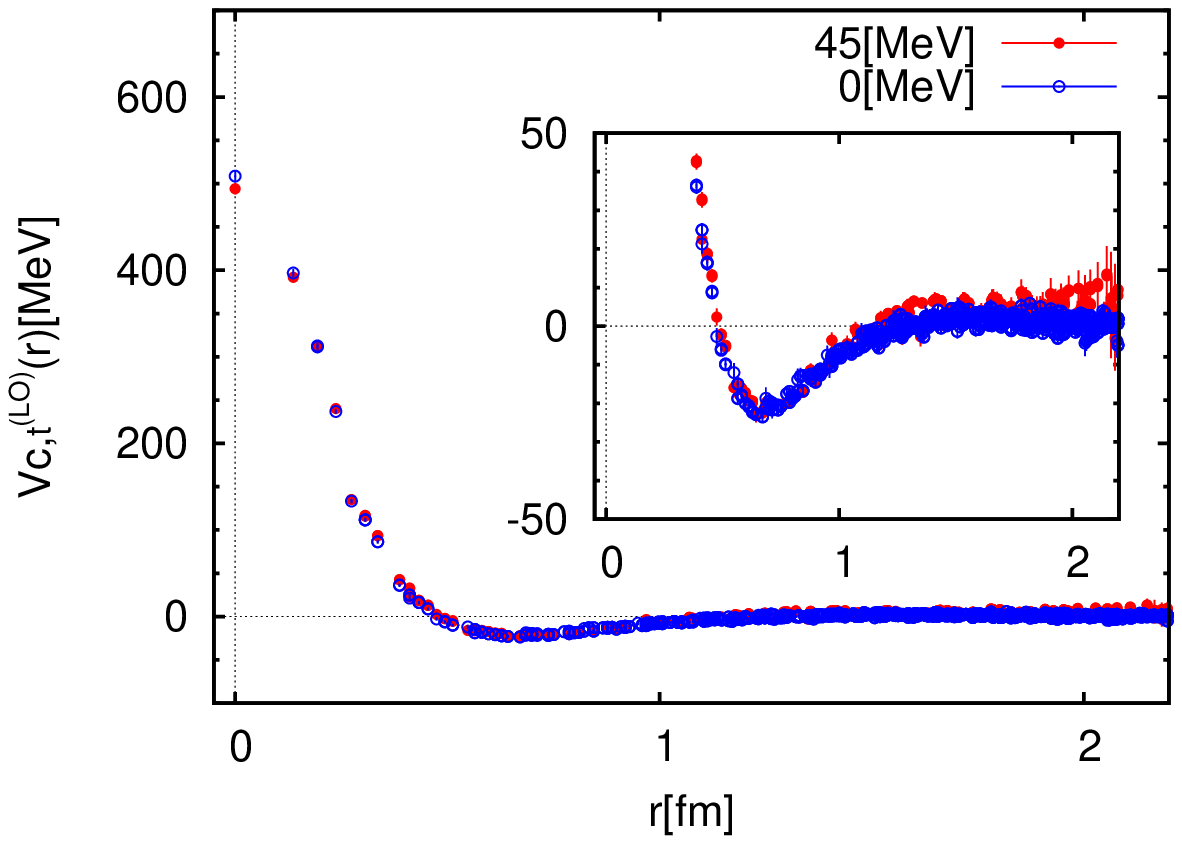}
\includegraphics[width=6.5cm]{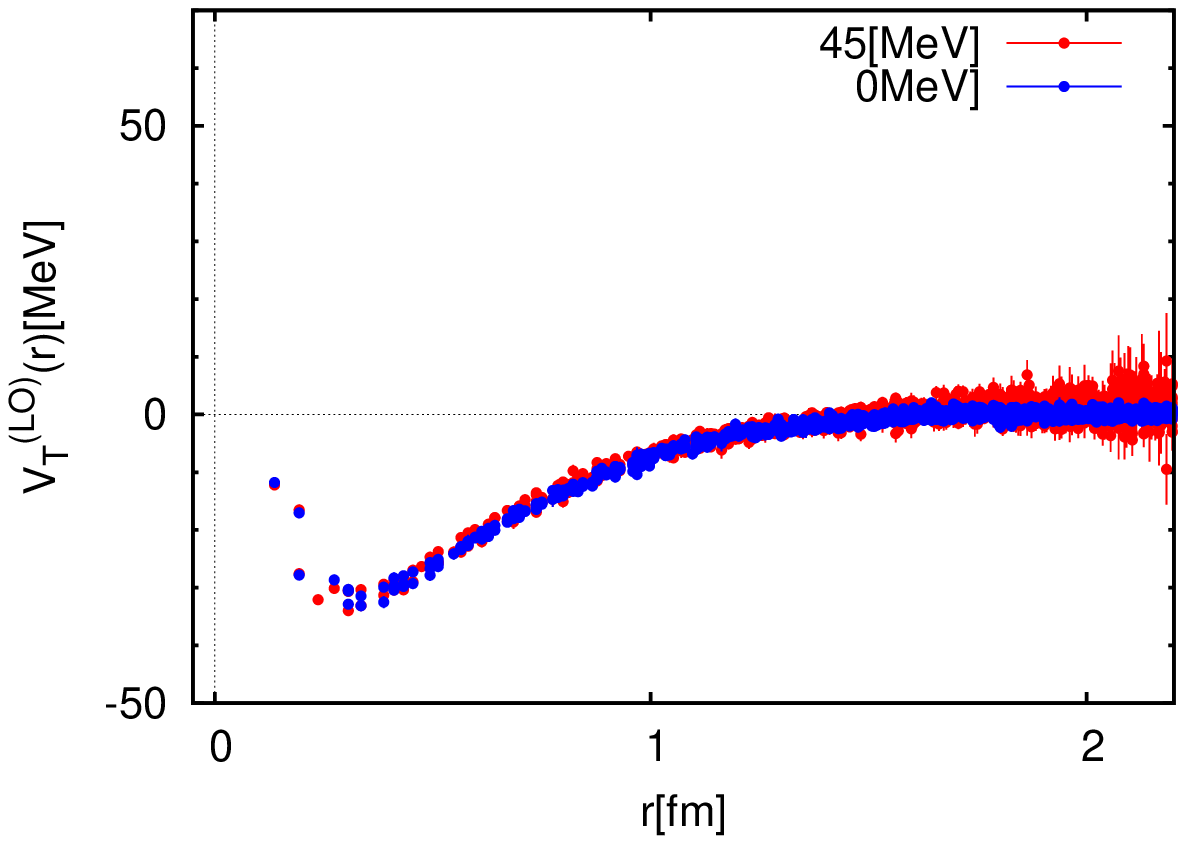}
\end{center}   
  \caption{ 
    (Left)  The  LO central potential $V_{\rm  C,t}^{\rm (LO)}(r)$  for the
 spin-triplet and  the orbital
 $^3\!\,{\rm S}_1 - ^3\!\,{\rm D}_1$
 coupled channel as a function of $r$. 
    (Right)  The  LO tensor potential  $V_{\rm  T}^{\rm (LO)}(r)$  for the
 spin-triplet and  the orbital
$^3\!\,{\rm S}_1 - ^3\!\,{\rm D}_1$
coupled channel as a function of $r$.
    Symbols are same as in \Fig{fig:singlet_APBCvsPBC}.
}
 \label{fig:triplet_APBCvsPBC}
\end{figure}

In \Fig{fig:singlet_3d}, we plot
 the spin-singlet central potentials $V_{\rm C,s}^{\rm LO}(x,y,z=0)$
 obtained from  the  corresponding
 NBS wave functions $\phi(x,y,z=0)$ in \Fig{fig:1S0wave_3d}.
  Although the wave functions have different spatial structure for
  different energies, the potentials are independent of $E$ and localized in space. 

 To make more  precise comparison, $V_{\rm C,s}^{\rm (LO)}(r)$ is plotted 
 as a function of $r$  for $E\simeq 0$ MeV (blue) and
 at  $E \simeq 45$ MeV (red) in \Fig{fig:singlet_APBCvsPBC}(Left).
 Similar comparisons are also made for $V_{\rm C,t}^{\rm (LO)}(r)$ 
 and $V_{\rm T}^{\rm (LO)}(r)$  in \Fig{fig:triplet_APBCvsPBC}.
 In all these cases, we find no $E$-dependence within statistical errors.
We therefore conclude  that the  LO potential is 
a good  approximation  for
$U(\br, \br^\prime)$ in the energy range $E=0$ -- $45$ MeV.

It  should be kept  in mind  that we employ  a large  pion mass
$m_{\pi} \simeq  0.53$ GeV, which  may be one  of the reasons  for the
small  energy dependence  of  the LO  potentials.   It is  therefore
important to increase $E$ or decrease $m_{\pi}$ and check the point where
 NLO contributions become significant. 

\subsection{LO  potentials for different orbital angular momentum}

As mentioned in Sec.\ref{sec:boundary},
 source functions in \Eq{eq:source_APBC} for the APBC generate not only the orbital $A_1^+$ but also the orbital $T_2^+$ components. Combining these sources appropriately,
one can construct the NBS wave function for the spin-singlet and the orbital $T_2^+$ channel
($\simeq ^1\!{\rm D}_2$ state). Therefore  
the LO central potential $V_{\rm C,s}^{\rm (LO)}(r)$ can be
extracted also from this wave function.

In \Fig{fig:singlet_APBCvsPBC}(Right), 
$V_{\rm C,s}^{\rm (LO)}(r)$ obtained from the orbital $T_2^+$ channel is compared with the same potential determined from the orbital $A_1^+$ channel at $E\simeq 45$ MeV. Although statistical errors are large,  we observe that the
 two determinations give consistent result.
Assuming that the orbital $A_1^+$ and $T_2^+$ representations
 are dominated by $\ell=0$ and $\ell=2$ waves, respectively, we here conclude that the LO potential in  the derivative expansion  is a
good  approximation of  $U(\br, \br^\prime)$ for  $\ell\le 2$ in the spin-singlet and 
 positive parity channel.
 
\section{Summary and conclusion}
\label{summary}
 We have studied the validity of derivative expansion of the 
  energy-independent non-local NN potential 
  $U(\br,\br')=  V(\br, {\nabla}_{\brs})  \delta^3(\br-\br^\prime)$, 
  defined from the  NBS wave function on the lattice.
 For this purpose,  we have carried out quenched lattice QCD simulations 
  for the NBS wave functions with the lattice spacing 0.14 fm,
  spatial lattice size 4.4 fm and the pion mass
   $m_{\pi}\simeq 530$  MeV. Relative kinetic  energies between two nucleons were
  controlled by employing the periodic and anti-periodic boundary conditions for the 
   quark fields in the spatial directions.  

  The leading-order  potentials   obtained 
  at different energies ($E \simeq  0$ MeV and $45$ MeV) show
  no difference within statistical
  errors, which validates the local approximation of the potential up to $E=45$ MeV 
  for the central and tensor potentials.  
  We have also compared the central potentials in the spin-singlet channel
  for different orbital angular momentum ($\ell=0$ and $\ell=2$)
  at $E \simeq 45$ MeV. The result also supports the validity of 
   the local approximation of the potential at this energy. 
 
In the  future it is important to  apply the analysis
  in this report to the general baryon-baryon potentials 
  in full QCD for lighter quark masses  (smaller inelastic threshold  $E_{\rm th}$) 
  and for smaller lattice spacing $a$
   to investigate the convergence of the 
   derivative expansion in realistic situations.

 \section*{Acknowledgments}
 We are grateful for authors and maintainers of {\tt CPS++}\cite{cps},
 a  modified version of  which is  used for  simulations done  in this
 report.  This work is supported by the Large Scale Simulation Program
 No.08-19(FY2008)  and  No.09-23(FY2009)  of High  Energy  Accelerator
 Research Organization (KEK).  This work  was supported in part by the
 Grant-in-Aid of  the Ministry  of Education, Science  and Technology,
 Sports and Culture (Nos. 20340047, 20105001, 20105003, 22540268) and
in part by a Grand-in-Aid for Specially Promoted Research (13002001).

\vskip 1.0cm

\appendix
\section{Spatial rotation on the lattice}
\label{sec:lattice_sym}
\subsection{Cubic group ${\rm SO}(3,{\mathbb{Z}})$}
\label{sec:cubic}

In this Appendix, we present a brief summary on the symmetry of 
 two nucleon system on the lattice.
 Further account on the representations of the cubic
group can be seen in Refs.\citen{Wigner,Landau}.
Note that the NBS wave functions in higher partial waves on the lattice  are
first discussed by L\"uscher \cite{Luscher:1990ux}.

A relation of irreducible representations between ${\rm SO}(3,{\mathbb{Z}})$ and  ${\rm SO}(3,{\mathbb{R}})$ is given in table~\ref{tab:cubic} for $\ell \le 6$. For two nucleon, the total spin $S$ becomes $1/2\otimes 1/2 = 1 \oplus 0$, which corresponds to $T_1$($S=1$) and $A_1$($S=0$) of the ${\rm SO}(3,{\mathbb{Z}})$.
Therefore, the total representation $J$ for two nucleon system is determined by the product $J=R_1\otimes R_2$, where $R_1=A_1,A_2,E,T_1,T_2$ for the orbital "angular momentum" while $R_2=A_1, T_1$ for the total spin. In table~\ref{tab:product}, the product $R_1\otimes R_2$ is decomposed into the direct sum of irreducible representations. 
For example, if the two nucleon state in the spin-triplet ($R_2=T_1$) belongs to the $J^{P}=T_1^+$ representation, the orbital representation $R_1$ should satisfy $T_1^+ \in (R_1 \otimes T_1)$. From the table~\ref{tab:product}, solutions to this condition are given by $R_1=A_1^+$, $E^+$, $T_1^+$ and $T_2^+$.   

\begin{table}[h]
\begin{center}
\caption{Numbers of each representation of  ${\rm SO}(3,{\mathbb{Z}})$ which appears in the angular momentum $\ell$ representation of ${\rm SO}(3,{\mathbb{R}})$. $P=(-1)^\ell$ represents an eigenvalue under parity transformation.}
\label{tab:cubic}
\vspace{0.3cm}
\begin{tabular}{|cc|ccccc|}
\hline
$\ell$ & $P$ & $A_1$ & $A_2$ & $E$ & $T_1$ & $T_2$ \\
\hline
0 (S)&  $+$ & 1 & 0 &0 &0 & 0 \\
1 (P)&  $-$  & 0 & 0 &0 & 1 & 0\\
2 (D)&  $+$ & 0 & 0 &1 & 0 & 1 \\
3 (F)&  $-$  & 0 & 1 &0 & 1 & 1\\
4 (G)&  $+$ & 1 & 0 &1 & 1 & 1 \\
5 (H)&  $-$  & 0 & 0 &1 & 2 & 1\\
6 (I) &  $+$ & 1 & 1 &1 & 1 & 2 \\
\hline
\end{tabular}
\end{center}
\end{table}     

\begin{table}[h]
\begin{center}
\caption{A decomposition for a product of two irreducible representations, $R_1\otimes R_2$, into irreducible representations in ${\rm SO}(3,{\mathbb{Z}})$.  Note that $R_1\otimes R_2= R_2\otimes R_1$ by definition. }
\label{tab:product}
\vspace{0.3cm}
\begin{tabular}{|c||ccccc|}
\hline
  & $A_1$ & $A_2$ & $E$ & $T_1$ & $T_2$ \\
\hline
\hline
$A_1$ & $A_1$ & $A_2$ & $E$ & $T_1$ & $T_2$ \\
$A_2$   & $A_2$  & $A_1$ & $E$ & $T_2$ &  $T_1$\\
$E$       &  $E$ & $E$  &$A_1\oplus A_2\oplus E$ & $T_1\oplus T_2$ & $T_1\oplus T_2$ \\
$T_1$    & $T_1$ & $T_2$& $T_1\oplus T_2$ & $A_1\oplus E \oplus T_1\oplus T_2$ & $A_2\oplus E \oplus T_1\oplus T_2$\\
$T_2$  & $T_2$ & $T_1$  &$T_1\oplus T_2$& $A_2\oplus E \oplus T_1\oplus T_2$ &$A_1\oplus E \oplus T_1\oplus T_2$  \\
\hline
\end{tabular}
\end{center}
\end{table} 
   
\subsection{The cyclic group $C_4$}
\label{sec:c4}

Elements of SO(3, ${\mathbb{Z}}$)  which correspond to the rotation around
the z-axis form a cyclic group $C_4$ consisting of four elements
\begin{equation}
  C_4
  \equiv
  \{
  e, c_4, (c_4)^2, (c_4)^3
  \},
\end{equation}
where $e$ denotes the identity,  and $c_4$ denotes the rotation around
the z-axis by 90 degrees.
It  has four one-dimensional  irreducible representations,  i.e., $A$,
$B$,  $E_1$ and  $E_2$.
They are related to the irreducible representations of SO$(2,{\mathbb{R}})$
labeled by $M=0, 2, +1, -1$ (modulo 4), respectively.
The representation matrices are summarized in \Table{table:c4.rep}.
A  relation of  irreducible  representations between  $C_4$ and  SO(3,
${\mathbb{Z}}$) is given in \Table{table:c4.and.o}.
\begin{table}[h]
  \begin{center}
    \caption{Representation matrices of irreducible representations of $C_4$.}
    \label{table:c4.rep}
    \begin{tabular}{|c|cccc|}
      \hline
          & $e$ & $c_4$ & $(c_4)^2$ & $(c_4)^3$ \\
      \hline
      $A$   & 1 & 1     & 1         & 1         \\
      $B$   & 1 & $-1$  & 1         & $-1$      \\
      $E_1$ & 1 & $i$   & $-1$      & $-i$      \\
      $E_2$ & 1 & $-i$  & $-1$      & $i$       \\
      \hline
    \end{tabular}
  \end{center}
\end{table}
\begin{table}[h]
  \caption{Numbers of each representation  of $C_4$ which  appear in
    each representation of SO(3, ${\mathbb{Z}}$).}
  \label{table:c4.and.o}
  \begin{center}
    \begin{tabular}{|c||cccc|}
      \hline
           & $A$ & $B$ & $E_1$ & $E_2$ \\
      \hline\hline
      $A_1$ & 1 & 0 & 0 & 0 \\
      $A_2$ & 0 & 1 & 0 & 0 \\
      $E$   & 1 & 1 & 0 & 0 \\
      $T_1$ & 1 & 0 & 1 & 1 \\
      $T_2$ & 0 & 1 & 1 & 1 \\
      \hline
    \end{tabular}
  \end{center}
\end{table}

\section{A source operator for the $^1\!\,{\rm D}_2$ state} 
\label{sec:D-wave-op}

A  general projection  formula for  the  source operator  in the  spin
singlet sector is given by
\begin{equation}
  \mathcal{J}^{S=0; \Gamma, M}(f^{(0)})
  \equiv
  \frac{d_{\Gamma}}{24}
  \sum_{{\cal R}\in {\rm SO}(3,\mathbb{Z})}
  D^{(\Gamma)*}_{MM}({\cal R})
  \mathcal{J}_{\alpha\beta}({\cal R}^{-1}\circ f^{(0)})
  \cdot
  P^{(S=0,S_z=0)}_{\alpha\beta},
  \label{eq:projection.s=0.general}
\end{equation}
where $\Gamma$ labels the cubic group representations ($A_1,A_2,
T_1,  T_2, E$),   
$d_{\Gamma}$  denotes the  dimension of  the  representation $\Gamma$,
$D^{(\Gamma)}({\cal  R})$  denotes the  representation  matrix of  the
representation $\Gamma$, and
$M$ is a  label of the irreducible representations  of $C_4$, 
contained  in the  irreducible representation  $\Gamma$ of {\rm SO}(3,$\mathbb{Z}$).  (See
\Table{table:c4.and.o}.)
This $M$ corresponds  to the azimuthal quantum number up to modulo 4 due  to the cubic symmetry.

To  derive \Eq{eq:source_1d2}, we  consider the  subgroup $H$  in SO(3,$\mathbb{Z}$),
which leaves $f^{(0)}(\br)$ invariant
\begin{equation}
  H \equiv \{{\cal R}\in {\rm SO}(3,\mathbb{Z})| {\cal R}\circ f^{(0)} = f^{(0)}\}.
\end{equation}
$H$ is  generated by $c_2$,  which corresponds to the  rotation around
${\bf  m}\equiv   (1,-1,0)$  by  180  degrees,  and   by  $c_3$,  which
corresponds  to the  rotation around  ${\bf n}\equiv  (1,1,1)$  by 120
degrees.   $H$ consists of  six elements,  i.e., $H  \equiv \{  e, c_3,
(c_3)^2, c_2, c_2c_3, c_2(c_3)^2  \}$, where $e$ denotes the identity.
We decompose  SO(3,$\mathbb{Z}$) by right  cosets of $H$  as ${\rm SO}(3,\mathbb{Z}) =  \bigcup_{c\in C_4}
Hc$, where $Hc\equiv \{hc| h \in  H\}$ denotes a right coset of $H$ in
$G$, which can be labeled by elements of $C_4$.
We  use   the  coset  decomposition   to  arrange  the   summation  in
\Eq{eq:projection.s=0.general} as
\begin{eqnarray}
  \lefteqn{
    \mathcal{J}^{S=0; \Gamma, M}(f^{(0)})
  }
  \nonumber\\
  &=&
  \frac{d_{\Gamma}}{24}
  \sum_{c\in C_4}
  \sum_{h\in H}
  D^{(\Gamma)*}_{MM}(hc)
  \mathcal{J}_{\alpha\beta}((hc)^{-1}\circ f^{(0)})
  \cdot
  P^{(S=0,S_z=0)}_{\alpha\beta}
  \nonumber\\
  &=&
  \sum_{M'}
  \frac{d_{\Gamma}}{6}
  \sum_{h\in H}
  D^{(\Gamma)*}_{MM'}(h)
  \cdot
  \frac1{4}
  \sum_{c\in C_4}
  D^{(\Gamma)^*}_{M'M}(c)
  \mathcal{J}_{\alpha\beta}(c^{-1}\circ f^{(0)})
  \cdot
  P^{(S=0,S_z=0)}_{\alpha\beta},
\end{eqnarray}
where we used $h^{-1}\circ f^{(0)} = f^{(0)}$.
\Eq{eq:source_1d2} is  arrived at by noting the  following three facts
(i) $D^{(\Gamma)^*}_{M'M}(c)$  is diagonal,  which reduces to  a phase
factor  $e^{iMj\pi/2} \delta_{M'M}$,  (ii) $f^{(j)}  = (c_4)^{-j}\circ
f^{(0)}$, where  $c_4$ denotes  the rotation around  the z-axis  by 90
degrees, (iii) $\frac{d_{\Gamma}}{6} \sum_{h\in H}D^{(\Gamma)*}_{MM}(h)
= 1$ for $\Gamma = A_1$ and $T_2$.

\vskip 1.0cm
 


\begin{thebibliography}{99}
  \bibitem{Machleidt:2000ge}
	  R.~Machleidt,
	  Phys.\ Rev.\  C {\bf 63}, 024001 (2001)
	  %
  \bibitem{Wiringa:1994wb}
	  R.~B.~Wiringa, V.~G.~J.~Stoks and R.~Schiavilla,
	  Phys.\ Rev.\  C {\bf 51}, 38 (1995)
	  %
  \bibitem{Stoks:1994wp}
	  V.~G.~J.~Stoks, R.~A.~M.~Klomp, C.~P.~F.~Terheggen and J.~J.~de Swart,
	  Phys.\ Rev.\  C {\bf 49}, 2950 (1994)
	  %
        \bibitem{Ishii:2006ec}
          N.~Ishii, S.~Aoki and T.~Hatsuda,
          Phys.\ Rev.\ Lett.\  {\bf 99}, 022001 (2007)
          [arXiv:nucl-th/0611096].
          %
        \bibitem{Aoki:2009ji}
          S.~Aoki, T.~Hatsuda and N.~Ishii,
          Prog.\ Theor.\ Phys.\  {\bf 123}, 89 (2010)
          [arXiv:0909.5585 [hep-lat]];
          %
	  Comput.\ Sci.\ Dis.\  {\bf 1}, 015009 (2008)
	  [arXiv:0805.2462 [hep-ph]].
	  %
\bibitem{Nemura:2008sp}
  H.~Nemura, N.~Ishii, S.~Aoki and T.~Hatsuda,
  Phys.\ Lett.\  B {\bf 673}, 136 (2009)
  [arXiv:0806.1094 [nucl-th]].

\bibitem{Inoue:2010hs}
  T.~Inoue {\it et al.}  [HAL QCD collaboration],
  Prog.\ Theor. \ Phys.\ {\bf 124}, 591 (2010)
  [arXiv:1007.3559 [hep-lat]].

\bibitem{Inoue:2010es}
  T.~Inoue {\it et al.}  [HAL QCD Collaboration],
  arXiv:1012.5928 [hep-lat].

\bibitem{Doi:2010yh}
  T.~Doi for HAL QCD Collaboration,
  arXiv:1011.0657 [hep-lat].
 
\cite{Ikeda:2010sg}
\bibitem{Ikeda:2010sg}
  Y.~Ikeda {\it et al.} [HAL QCD Collaboration],
    arXiv:1002.2309 [hep-lat].
%
\bibitem{Kawanai:2010ev}
  T.~Kawanai and S.~Sasaki,
  Phys.\ Rev.\  {\bf D82}, 091501 (2010)
  [arXiv:1009.3332 [hep-lat]].
%
  \bibitem{Aoki:2008yw}
	  S.~Aoki, J.~Balog and P.~Weisz,
	  Prog.\ Theor.\ Phys.\  {\bf 121}, 1003 (2009)
	  [arXiv:0805.3098 [hep-th]].
	  %
\bibitem{Murano:2010hh}
  K.~Murano, N.~Ishii, S.~Aoki and T.~Hatsuda,
  PoS {\bf LATTICE2010}, 150 (2010)
  [arXiv:1012.3814 [hep-lat]].

\bibitem{Murano:2010tc}
  K.~Murano, N.~Ishii, S.~Aoki and T.~Hatsuda,
  PoS {\bf LAT2009}, 126 (2009)
  [arXiv:1003.0530 [hep-lat]].

\bibitem{Luscher:1986pf}
  M.~Luscher,
  Commun.\ Math.\ Phys.\  {\bf 105}, 153 (1986).

\bibitem{Luscher:1990ux}
  M.~Luscher,
  Nucl.\ Phys.\  B {\bf 354}, 531 (1991).

\bibitem{Lin:2001fi}
  C.~J.~D.~Lin, G.~Martinelli, C.~T.~Sachrajda and M.~Testa,
  Nucl.\ Phys.\ Proc.\ Suppl.\  {\bf 109A}, 218 (2002)
  [arXiv:hep-lat/0111033];

%
\bibitem{Aoki:2005uf}
  S.~Aoki {\it et al.}  [CP-PACS Collaboration],
  Phys.\ Rev.\  D {\bf 71}, 094504 (2005)
  [arXiv:hep-lat/0503025].

\bibitem{Ishizuka:2009bx}
  N.~Ishizuka,
  PoS {\bf LAT2009}, 119 (2009)
  [arXiv:0910.2772 [hep-lat]].

\bibitem{KR56}
W. Kr\'olikowski and J. Rzewuski, Nuovo Cimento, {\bf 4}, 1212 (1956).  

  \bibitem{okubo} 
          S.~Okubo and R.E.~Marshak,
          Ann.\ Phys.(NY) {\bf 4}, 166 (1958)
	  %
\bibitem{TW67}
R. Tamagaki and W. Watari, Prog. Theor. Phys. Suppl. No.~{\bf 39}, 23 (1967).
 
\bibitem{Fukugita:1994ve} 
	  M.~Fukugita, Y.~Kuramashi, M.~Okawa, H.~Mino and A.~Ukawa,
	  Phys.\ Rev.\  D {\bf 52}, 3003 (1995)
	  %
 
 \bibitem{t-dep}
 N.~Ishii {\it et al.} [HAL QCD Collaboration], in preparation. 
 
  \bibitem{cps}
	 CPS++
	  {\verb|http://qcdoc.phys.columbia.edu/chuiwoo_index.html|}(maintainer:
	  Chulwoo Jung).

\bibitem{Wigner}
   E.\ P.\ Wigner, ``Group Theory and Its Application to the Quantum
	Mechanics of Atomic Spectra,'' expanded and improved ed.  
	(Academic Press, New York, 1971).
	
\bibitem{Landau}
L.\ D.\ Landau, L.\ M.\ Lifshitz, ``Quantum Mechanics (Non-relativistic Theory)'',
 3rd ed. (Butterworth-Heinemann, New York, 1977).
 \end{thebibliography}
\end{document}